\documentclass[prd, reprint, amsmath,amssymb,nofootinbib,floatfix, aps, longbibliography, superscriptaddress, preprintnumbers]{revtex4-2}
\usepackage{graphicx}
\usepackage{bm}
\usepackage{graphicx}
\usepackage{dcolumn}
\usepackage{bbold}
\usepackage{mathtools}
\usepackage{siunitx}
\usepackage[colorlinks=true,urlcolor=rossos,anchorcolor=black,citecolor=mygreen,filecolor=black,linkcolor=black,menucolor=blue,linktocpage=true, linkcolor=blue]{hyperref}
\usepackage{booktabs}
\usepackage{colortbl}
\usepackage{enumerate}
\usepackage[normalem]{ulem}
\usepackage{float}
\usepackage{todonotes}
\usepackage{verbatim}
\usepackage{multirow}
\usepackage{xparse}
\usepackage{tabularray}
\UseTblrLibrary{booktabs,siunitx}
\usepackage{relsize}
\usepackage{xcolor}
\usepackage{wrapfig}
\usepackage{float}
\usepackage{siunitx} 
\usepackage{booktabs}

\DeclareSIUnit\clight{\text{\ensuremath{c}}}
\DeclareSIUnit\tevm{\TeV\per\clight\squared}
\DeclareSIUnit\events{events}
\DeclareSIUnit\years{y}
\DeclareSIUnit\tonne{t}
\DeclareSIUnit\tonneyears{\tonne\years}
\DeclareSIUnit\cmtwo{\cm\squared}
\DeclareSIUnit\kev{\kilo\eV} 
\DeclareSIUnit\mev{\mega\eV} 
\DeclareSIUnit{\GeV}{\giga\eV}
\DeclareSIUnit{\TeV}{\tera\eV}
\DeclareSIUnit{\kevm}{\keV\per\clight\squared}
\DeclareSIUnit{\mevm}{\MeV} 
\DeclareSIUnit{\gevm}{\GeV}
\DeclareSIUnit{\tevm}{\TeV}
\DeclareSIUnit\kevnr{\si{\kev} {}_\mathrm{NR}} 
\DeclareSIUnit\kever{\si{\kev} {}_\mathrm{ER}} 
\DeclareSIUnit\PE{\mathrm{PE}}
\DeclareSIUnit\bar{bar}

\definecolor{rossos}{cmyk}{0,1,1,0.55}

\definecolor{purple}{rgb}{0.7,0,1}
\definecolor{teal}{rgb}{0,0.5,0.5}

\definecolor{mygreen}{rgb}{0.2, 0.64, 0.48}
\definecolor{mygray}{gray}{.95}

\def\max{{\mathrm{max}}}

\usepackage{circuitikz}

\usepackage{hyperref}

\usepackage{mhchem}

\usepackage{placeins}

\begin{document}


\title{A low-cost compact muon detector for millikelvin operation}

\author{Tengiz Ibrayev}
\homepage{tengiz.ibrayev@sydney.edu.au}
\affiliation{The School of Physics, The University of Sydney, Camperdown, Sydney, Australia}

\author{Danish Khan}
\affiliation{The School of Physics, The University of Sydney, Camperdown, Sydney, Australia}

\author{Alex Herscovics}
\affiliation{The School of Physics, The University of Sydney, Camperdown, Sydney, Australia}
 
\author{Komal Kataruka}
\affiliation{The School of Physics, The University of Sydney, Camperdown, Sydney, Australia}

\author{Matilda Lawton}
\affiliation{The School of Physics, The University of Sydney, Camperdown, Sydney, Australia}

\author{Sandhya Sharma}
\affiliation{The School of Physics, The University of Sydney, Camperdown, Sydney, Australia}

\author{Steven Waddy}
\affiliation{The School of Physics, The University of Sydney, Camperdown, Sydney, Australia}
\affiliation{Emergence Quantum Pty. Ltd., Sydney, NSW, Australia.}

\author{Laura Manenti}
\homepage{laura.manenti@sydney.edu.au}
\affiliation{The School of Physics, The University of Sydney, Camperdown, Sydney, Australia}

\date{\today}

\begin{abstract}
Ionizing radiation produces correlated errors in superconducting qubits, posing some challenges to quantum error correction. In addition, it causes background events in other superconducting cryogenic devices, such as kinetic inductance detectors and superconducting nanowire single-photon detectors. 
Cosmic muons are a component of this background that can only be partly shielded against at ground level, but can be tagged by an independent detector inside the cryostat, so that the affected data can be vetoed, or the muon tags incorporated into error correction.

We present a low-cost, compact muon detector consisting of two plastic scintillator modules, each read out by a silicon photomultiplier (SiPM), operated in coincidence. We characterize the SiPMs from room temperature down to \qty{10}{\milli\kelvin}, including the breakdown and upper-limit voltages that define the operating bias range. 
Operated as a muon counter, the detector measures a coincidence rate that, as expected, is independent of temperature within uncertainties and agrees with analytical and Monte Carlo predictions.

Below approximately \qty{8}{\kelvin}, each muon event produces two pulses, one narrow and one broad, an effect that, to our knowledge, has not been reported before. Coincidences with an independent muon detector outside the cryostat confirm that muons are still counted reliably in this regime.
By studying the appearance of double pulses as a function of temperature and magnetic field up to \qty{4.5}{\tesla}, we find a behavior consistent with the phase diagram of a type-II superconductor. This links the double pulses to a superconducting material in the SiPM, likely an adhesion layer or diffusion barrier introduced during manufacturing. The mechanism by which the superconducting state produces the second pulse is not yet understood.
\end{abstract}
\maketitle

\section{Introduction \label{sec:intro}}
\begin{figure*}
    \centering
    \includegraphics[width=1\linewidth]{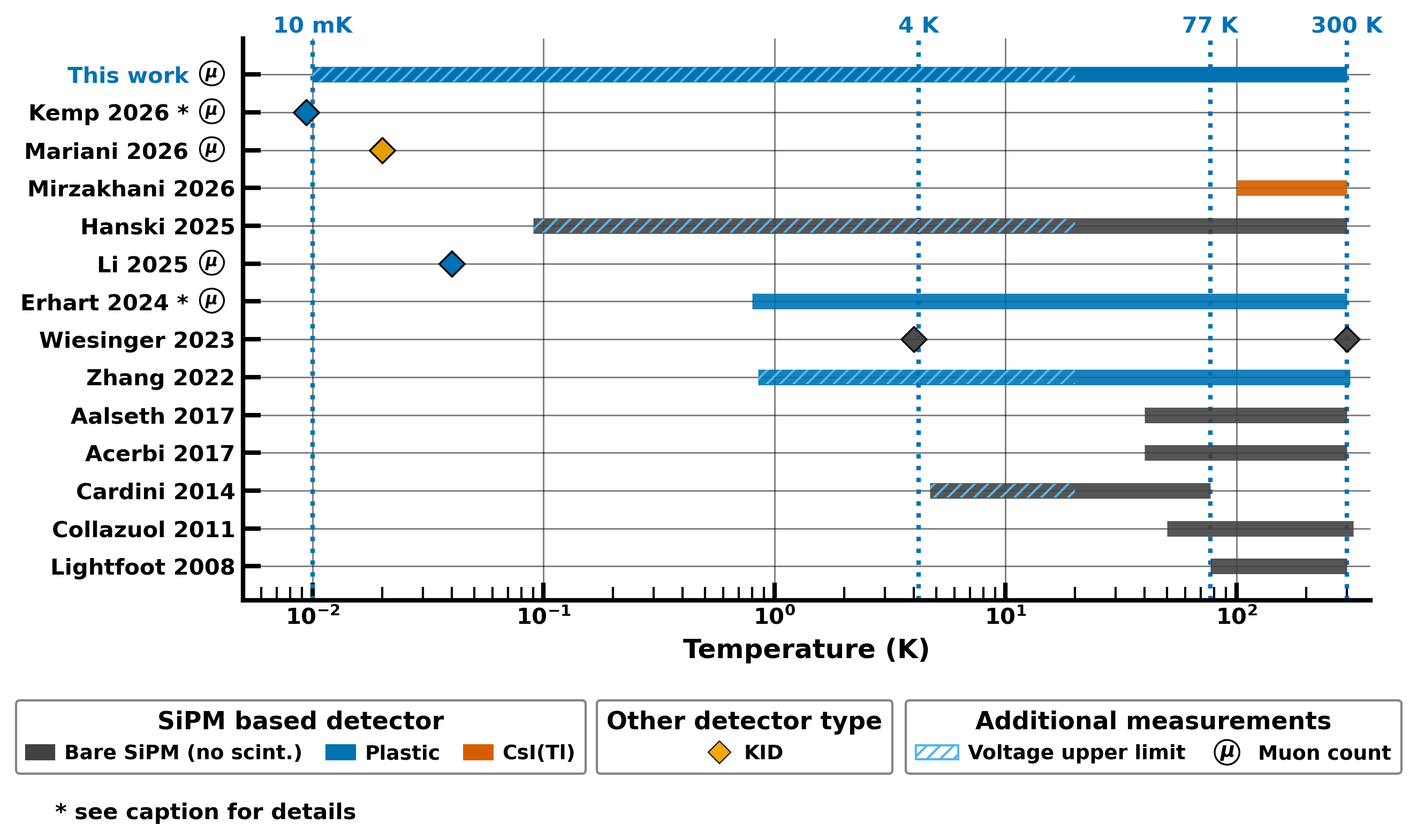}
    \caption{Summary of previous studies of SiPMs, scintillator–SiPM detectors and KIDs at cryogenic temperatures, together with this work. The y-axis lists the studies, and the x-axis shows the temperatures at which the detectors were operated. Bars indicate extended temperature ranges, while diamonds mark measurements at discrete temperatures. Colors identify the detector type and, where applicable, the scintillator material.  The hatching on the bars indicates the temperature ranges over which the upper-limit voltage, $V_{\mathrm{UL}}$, was measured (\autoref{sec:SiPM_intro}). The circled \textcircled{$\mu$} identifies studies that used coincidence measurements for muon counting. This work covers temperatures from 300 K to 10 mK and includes measurements of the breakdown voltage $V_{\mathrm{BD}}$ and the $V_{\mathrm{UL}}$, as well as muon counting. From left to right, the vertical dotted lines mark 10 mK, a typical dilution refrigerator base temperature; 4 K, approximately the boiling point of liquid helium; 77 K, approximately the boiling point of liquid nitrogen; and 300 K, room temperature. The works shown in the plot are, from top to bottom, those of \citet{Kemp2026}, \citet{Mariani2026}, \citet{Mirzakhani2026}, \citet{Hanski2025}, \citet{Li2025}, \citet{Erhart2024}, \citet{wiesinger2023}, \citet{Zhang2022}, \citet{Aalseth2017}, \citet{Acerbi2017}, \citet{Cardini2014}, \citet{Collazuol2011}, and \citet{Lightfoot2008}. In \citet{Erhart2024}, only the scintillator was cooled to cryogenic temperatures while the SiPMs were operated at room temperature. \citet{Kemp2026} develops a cryogenic scintillation detector capable of detecting muons, but does not measure a muon count rate.
}
    \label{fig:LITREVIEWPLOT}
\end{figure*}

Superconducting devices are a leading platform for quantum  computation~\cite{Kjaergaard2020}. Errors in individual physical qubits still limit how many operations can be performed reliably. Practical quantum computing therefore relies on quantum error correction (QEC), which encodes quantum information across many physical qubits to form a smaller number of robust logical qubits.

QEC can suppress errors only if the error rate per cycle stays below a threshold. This threshold is derived assuming that errors are independent and occur randomly in space and time. Errors caused by ionizing radiation violate both assumptions at once.

Energetic particles from environmental radioactivity and cosmic rays deposit energy in the device substrate. The substrate is much thicker than the superconducting
film and extends under the whole chip, so it absorbs most of this energy. The energy then spreads through the substrate as phonons. When the phonons reach the
superconducting film, they break Cooper pairs and create quasiparticles in many qubits at once. The quasiparticles shorten the energy-relaxation time $T_1$ of these qubits~\cite{Vepsalainen2020}, so their error rate rises sharply and stays high until the quasiparticles recombine. Each impact therefore produces errors on many qubits at once and over many consecutive QEC cycles: the errors are correlated in both space and time~\cite{Wilen2021,McEwen2022}.

Ionizing radiation interacting with the substrate affects other superconducting cryogenic devices: kinetic inductance detectors (KIDs), for which cosmic-ray glitches are a well-known problem~\cite{Karatsu2019}, and superconducting nanowire single-photon detectors, in which counts of unidentified origin limit dark matter searches~\cite{Chiles2022, QROCODILE2025}.
The difficulty in each case is distinguishing background events from the signal the device is built to measure. An independent detector can help identify events caused by external interactions.

The detector presented in this paper provides this independent tag for the cosmic muon component of the background. Muons dominate the charged particle flux at the Earth's surface, approximately 1~cm$^{-2}$~min$^{-1}$ at sea level~\cite{PDGCosmicRays2022}. They are far more penetrating than the $\alpha$, $\beta$ and $\gamma$ radiation from radioactive decay: a few centimeters of lead attenuate a MeV $\gamma$ ray by orders of magnitude, whereas the same shield stops only a small fraction of muons, and a substantial reduction of the muon flux requires going deep underground~\cite{Groom2001}.

Our detector consists of two modules, each comprising a plastic scintillator coupled to its own silicon photomultiplier (SiPM). SiPMs were chosen for their compact size, low cost, and commercial availability. The modules are operated in coincidence to identify traversing muons while rejecting uncorrelated backgrounds. The compact design allows the detector to be placed inside a cryostat, close to, and aligned with a superconducting device, to tag muons that cross its substrate. 

A detector of this kind has recently been shown to work: \citet{Li2025} operated two plastic scintillators with SiPM readout inside a dilution refrigerator housing a 63-qubit flip-chip processor and used their coincidence to demonstrate that a subset of the observed quasiparticle bursts were muon-induced. The detectors themselves, however, were not characterized and their quoted muon rate was not supported by a simulation of the detector acceptance. How SiPMs behave at millikelvin temperatures, and how a cryogenic detector can be built and calibrated to tag muons in the same cryostat as a superconducting device, have been addressed in separate studies belonging to communities that have so far made little reference to one another~\cite{Erhart2024, Hanski2025, Kemp2026, Mariani2026, wiesinger2023, Zhang2022, Mirzakhani2026, Li2025, Zhang2022, Aalseth2017, Acerbi2017, Cardini2014, Collazuol2011, Lightfoot2008}.

A picture is often worth a thousand words. We therefore bring the available literature together in \autoref{fig:LITREVIEWPLOT}, showing which detectors have been studied, at what temperatures, and what was measured. The overview reveals a gap: no study brings together SiPM characterization, measurements with the assembled scintillator modules, and muon coincidence counting across the full range from room temperature to millikelvin temperatures, with measured coincidence rates compared with Monte Carlo predictions.

In this work, we address this gap. We characterize the SiPMs from room temperature down to approximately 10~mK, measuring their breakdown and upper-limit voltages. We further characterize the voltage gain, optical crosstalk probability, relative photon detection efficiency, and single photoelectron (SPE) resolution down to approximately \qty{46}{K}. We then operate the assembled detector as a muon counter over the entire temperature range and show that the coincidence rate does not change with temperature and agrees with the Monte Carlo predictions. In future work, we plan to install the detector alongside a qubit array to tag muon interactions, allowing potentially affected data to be flagged or vetoed, and to investigate how these tags could inform error correction.

The paper is organized as follows. \autoref{sec:SiPM_intro} introduces SiPM operating principles. \autoref{sec:Experimental} describes the characterization setup, the detector, and muon rate data acquisition, and \autoref{sec:Results} presents the results. \autoref{sec:Double_pulses} discusses an unexpected observation: below approximately 8~K, each muon event produced two pulses rather than one. \autoref{sec:Conclusions} summarizes our conclusions.

\section{SiPM operating principles \label{sec:SiPM_intro}}
\begin{table*}[t]
    \caption{Cryostats used in this work, listed in order of decreasing base temperature.}
    \label{tab:cryostats}
    \begin{tblr}{width=\textwidth, colspec={llllX[l]}}
        \toprule
        Name & Manufacturer/Model & Type & {Base temp.} & Use in this work \\
        \midrule
        \emph{Attocube} & {attocube\\attoDRY800} & Closed-cycle cryostat & {\qty{4}{\kelvin}\\ (nominal)} & SiPM characterization \\
                \emph{Leiden-4K} & {Leiden Cryogenics\\CF-CS81-4K} & \qty{4}{\kelvin} system & \qty{3.6}{\kelvin} & SiPM characterization, scintillator--SiPM module tests \\
        \emph{Leiden-mK} & {Leiden Cryogenics\\CF-450} & Dilution refrigerator & \qty{20}{\milli\kelvin} & Muon rate, SiPM characterization, scintillator--SiPM module tests \\
        \emph{Bluefors} & {Bluefors\\LD400} & Dilution refrigerator & \qty{10}{\milli\kelvin} & Muon rate, SiPM characterization, scintillator--SiPM module tests, tests in a \qty{4.5}{\tesla} magnetic field \\\bottomrule
    \end{tblr}
\end{table*}
A SiPM is a solid-state photodetector consisting of an array of several hundred to several tens of thousands of microcells. In this work, we used the Onsemi C-series MicroFC-60035, with an area of $6\times$\qty{6}{\milli\metre\squared} and 18,980 microcells~\cite{OnsemiMicroC2022}. Each microcell comprises a single-photon avalanche diode (SPAD) connected in series with a quenching resistor, $R_Q$, and all microcells are connected in parallel between a common anode and cathode.

\begin{figure}
    \centering
    \includegraphics[width=\linewidth]{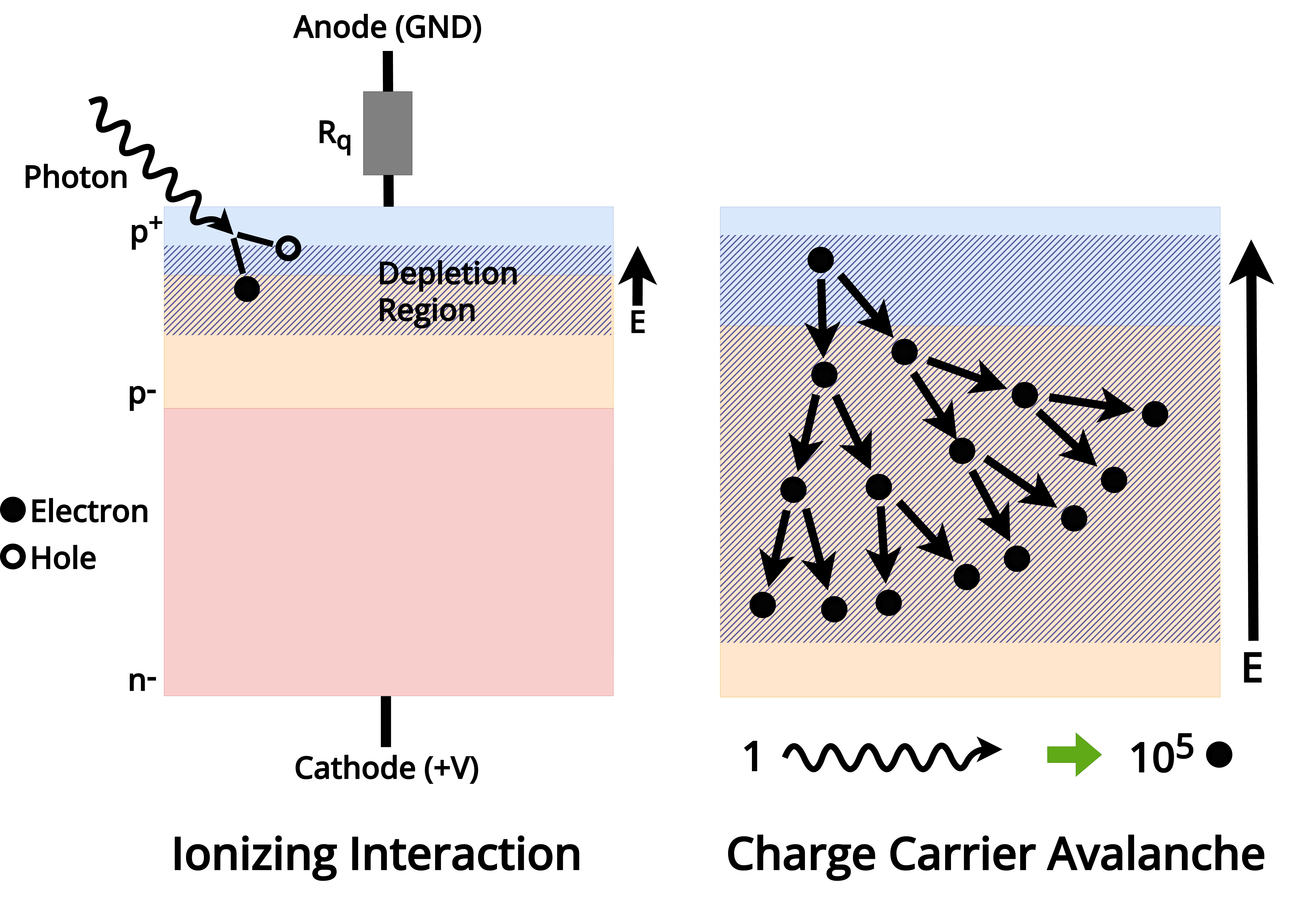}
    \caption{Operating principle of a SPAD, the main element of a SiPM microcell. Left: a free charge carrier is generated in the depletion region by an absorbed photon. The electric field $E$ in the depletion region separates the electron and the hole. Right: above the breakdown voltage, the accelerated carrier generates further electron--hole pairs through impact ionization, producing an avalanche of order $10^{5}$ carriers from a single initial carrier.}
    \label{fig:SiPM_avalanche}
\end{figure}

A SPAD is a reverse-biased $p$--$n$ junction operated above its breakdown voltage, $V_{\mathrm{BD}}$. A $p$--$n$ junction contains a depletion region, where mobile charge carriers have diffused away from the junction, leaving fixed ionized dopants that generate an electric field. When a free charge carrier is generated in or near this region, typically by the absorption of a photon, the electric field separates the electron and the hole (\autoref{fig:SiPM_avalanche}, left). Applying a reverse bias widens the depletion region and increases the electric field which can become sufficiently strong for charge carriers to produce further electron--hole pairs through impact ionization. When the bias exceeds  $V_{\mathrm{BD}}$, the field is strong enough for a single carrier to initiate a self-sustaining avalanche (\autoref{fig:SiPM_avalanche}, right) and the SPAD operates in Geiger mode: once an avalanche is triggered, the charge released by a microcell is independent of the number of carriers that initiated it. A SiPM measures the light intensity by summing the signals from many microcells. As long as two or more photons rarely hit the same microcell at the same time, the number of fired microcells, and hence the SiPM output, is proportional to the number of incident photons.

The avalanche is quenched by $R_Q$. Since the SPAD and $R_Q$ are connected in series, the voltage across the SPAD is
\begin{equation}
V_{\mathrm{SPAD}} = V_{\mathrm{bias}} - I R_Q .
\end{equation}
As the avalanche current $I$ increases, the voltage drop $IR_Q$ across the quenching resistor increases. Since the bias voltage applied to the whole microcell $V_{\mathrm{bias}}$ is fixed, $V_{\mathrm{SPAD}}$ falls. Once $V_{\mathrm{SPAD}}$ drops below $V_{\mathrm{BD}}$, the avalanche can no longer sustain itself and is quenched. The microcell then recharges through $R_Q$ and returns to its initial Geiger-mode state \cite{WhatSiPMHowa}.

The SiPM operating point is conventionally described by the overvoltage,
\begin{equation}
V_{\mathrm{OV}} = V_{\mathrm{bias}}-V_{\mathrm{BD}},
\label{eq:overvoltage}
\end{equation}
which controls the gain, and avalanche triggering probability, while also affecting the photon detection efficiency, dark count rate, afterpulsing and optical crosstalk between cells. Since $V_{\mathrm{BD}}$ depends on temperature, cryogenic operation requires determining $V_{\mathrm{BD}}(T)$ in order to maintain a fixed $V_{\mathrm{OV}}$ during cooldown.

At higher bias, a second sharp increase in current can be observed under dark or very low-light conditions. We define the voltage at which this occurs as the operational upper-limit voltage, $V_{\mathrm{UL}}$, above which the rapidly increasing dark current prevents normal SiPM operation. This second increase is not a breakdown of the junction; it is commonly attributed to correlated noise: afterpulses and crosstalk trigger further avalanches, until the probability of correlated noise approaches 100\% and the current diverges~\cite{darroch2025measuring}. The effect is reversible, but the large current dissipates power in the SiPM. At cryogenic temperatures, this heating can make the cryostat temperature unstable; at higher temperatures, it can damage the device. The useful bias range is therefore
\begin{equation}
V_{\mathrm{BD}} < V_{\mathrm{bias}} < V_{\mathrm{UL}}.
\end{equation}

Both $V_{\mathrm{BD}}$ and $V_{\mathrm{UL}}$ change with temperature, so the bias range available for detector operation changes during cooldown. To operate the detector reliably at any temperature, $V_{\mathrm{BD}}$ and $V_{\mathrm{UL}}$ must therefore be known as a function of $T$. Their temperature dependence is presented in \autoref{sec:VBD_VUL_vs_T}.

\begin{figure}
    \centering
    \includegraphics[width=1\linewidth]{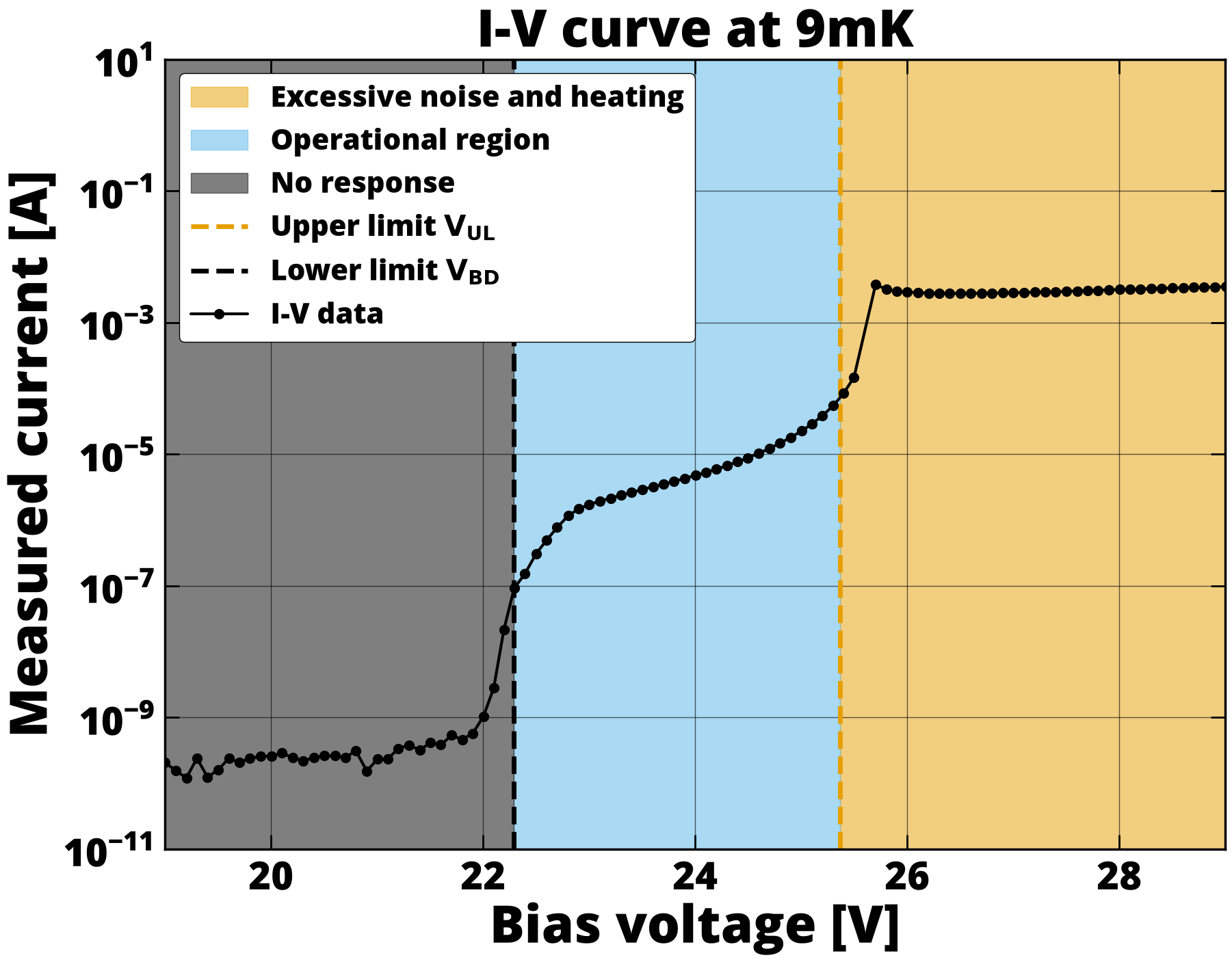}
    \caption{Current drawn by a SiPM as a function of bias voltage under dim blue LED illumination at \qty{9}{\milli\kelvin} (the curve is a linear interpolation of the data points). In the gray region, below the breakdown voltage $V_{\mathrm{BD}}$ (black dashed line), the current is small and the SiPM is not operating. The blue region indicates the operating range, where the photocurrent is proportional to the incident light. Above $V_{\mathrm{UL}}$ (orange dashed line), in the orange region, the current rises sharply until it reaches the compliance limit set on the SMU, resulting in excessive heating and unstable refrigerator operation.}
    \label{fig:I_vs_bias_VBD_VIIBD_example}
\end{figure}

\section{Experimental \label{sec:Experimental}}
\begin{figure*}
    \centering
         \includegraphics[height=0.3\textwidth]{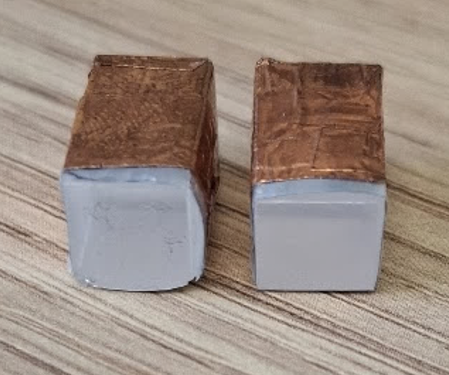}
         \includegraphics[height=0.3\textwidth]{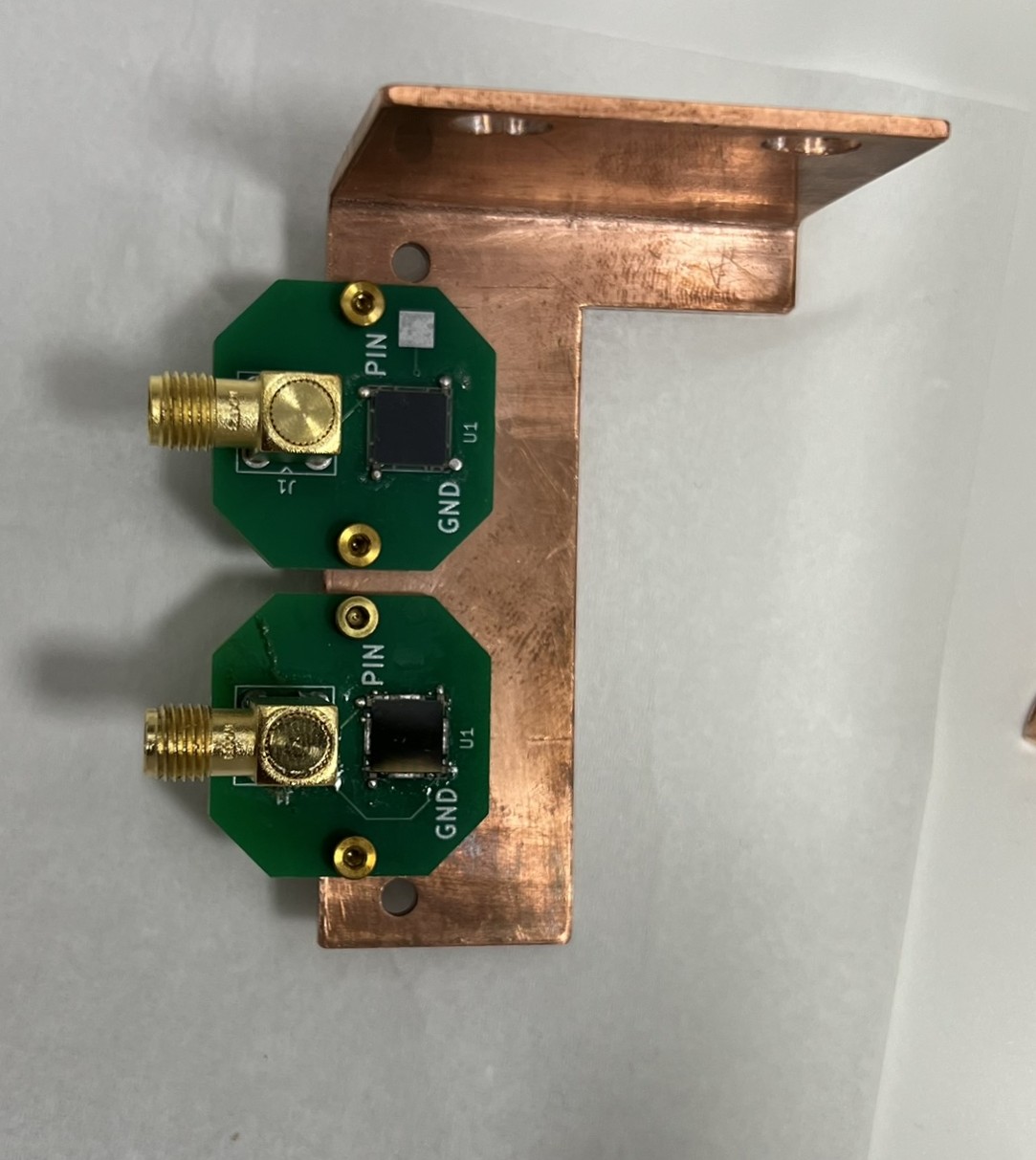}
         \includegraphics[height=0.3\textwidth]{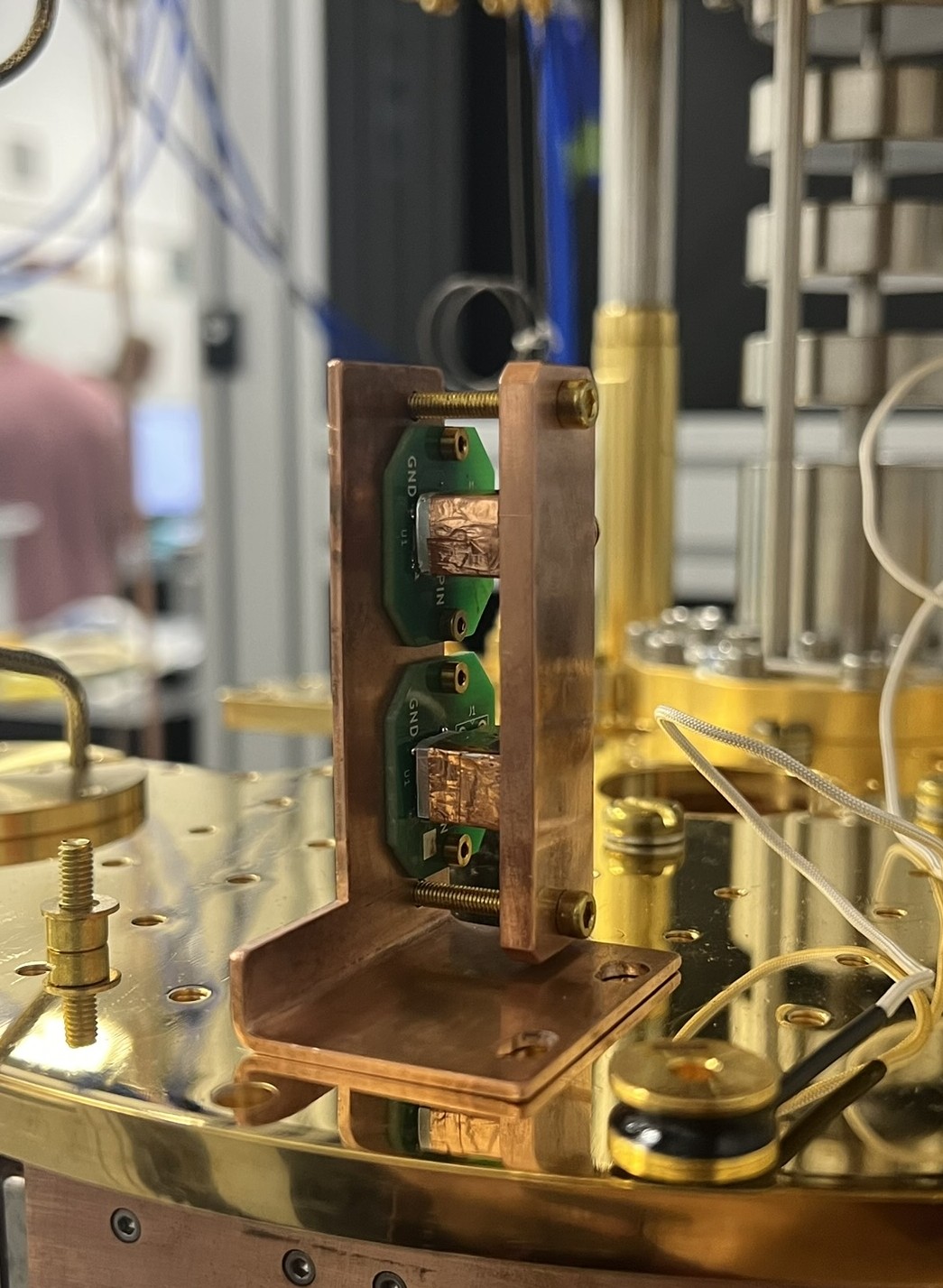}
    \caption{Photographs of the muon detector and its components. Left: Two plastic scintillators wrapped in ESR film and copper tape. Center: SiPMs mounted to PCBs screwed to a copper bracket. Right: The fully assembled detector, ready to be mounted to the mixing chamber of the dilution refrigerator.}
    \label{fig:reallifepictures}
\end{figure*}

\subsection{Cryostats}\label{sec:cryostats}
Cryogenic measurements were performed in the four cryostats listed in~\autoref{tab:cryostats}. In the rest of the paper, each cryostat is referred to by an abbreviated name. The breakdown and upper-limit voltage measurements are described in \autoref{sec:VBD_VUL_vs_T}, while the measurements of SiPM gain, photon detection efficiency, photon peak resolution and crosstalk probability are reported in Appendix~\ref{sec:SiPM_characterization}. The muon rate measurements are presented in \autoref{sec:muon_rate}. The measurements in a magnetic field were made to investigate the double pulses observed below \qty{8}{\kelvin} and are discussed in \autoref{sec:Double_pulses}.

\subsection{SiPM I--V measurements}\label{sec:IV_setup}
We determine $V_{\mathrm{BD}}$ and $V_{\mathrm{UL}}$ as a function of temperature from $I$--$V$ curves. We measured $I$--$V$ curves with a Keithley 2450 source measure unit (SMU) that supplied the bias voltage and measured the current.  Several methods can be used to extract the two voltages, as discussed in Appendix~\ref{app:vbdandul}. Here, we identify the breakdown voltage as the position of the corresponding maximum of $\mathrm{d}(\ln I)/\mathrm{d}V$. Figure~\ref{fig:I_vs_bias_VBD_VIIBD_example} shows an example of an $I$--$V$ curve collected under dim illumination, with both $V_{\mathrm{BD}}$ (first rise) and $V_{\mathrm{UL}}$ (second rise) visible.

A blue LED is used to induce a photocurrent, so that $V_{\mathrm{BD}}$ can be observed even in the absence of thermal dark counts. Conversely, $V_{\mathrm{UL}}$ is easier to measure with the SiPM in the dark: without light, the $I$--$V$ curve shows only one rise, and there is less risk of reaching the SMU current limit before a clear $V_{\mathrm{UL}}$ rise appears. $V_{\mathrm{UL}}$ is measured only below 40\,K; at higher temperatures, the required bias voltage would be too high for safe operation of the SiPM.

\begin{figure*}
    \centering
    \includegraphics[width=1\linewidth]{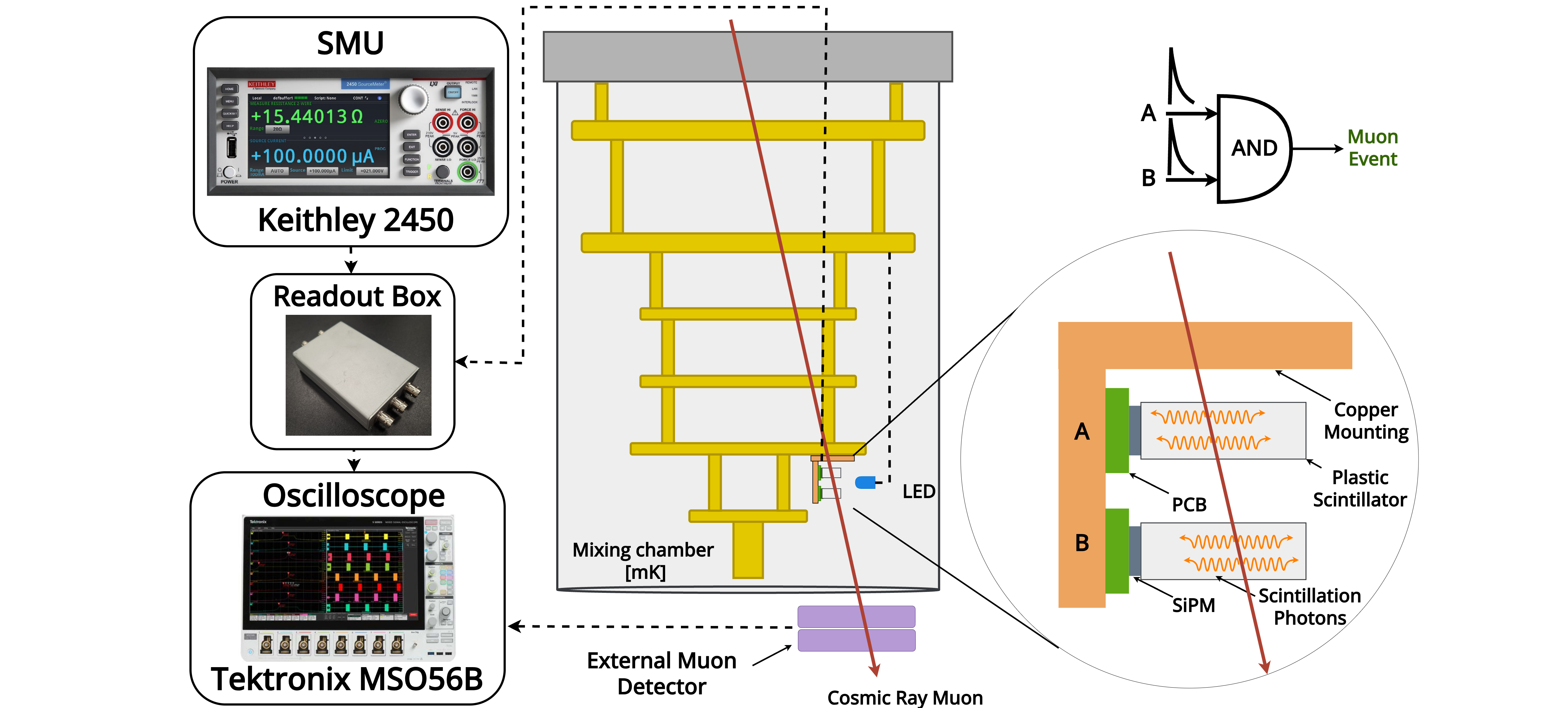}
    \caption{Experimental setup for the muon rate measurements. The Keithley 2450 SMU biases the SiPMs through the readout box, which also decouples the output signals, and the signals are recorded on the oscilloscope. The detector is mounted on the mixing chamber of the dilution refrigerator, and two external muon detectors are placed below the cryostat. Their signals are recorded together with those of the internal detector to identify quadruple coincidences and confirm that the internal coincidences are caused by muons. The inset shows the cryogenic muon detector, with the plastic scintillators coupled to SiPMs mounted on PCBs, and the AND logic used to trigger on coincidences. In some runs, additional bare SiPMs were mounted next to the detector with a blue LED on the \qty{4}{\kelvin} plate illuminating them for the $I$--$V$ characterization.}
    \label{fig:full_schematic}
\end{figure*}

\subsection{Muon detector design and assembly}\label{sec:detector_assembly}
The muon detector is made up of two identical modules, each comprising a SiPM and a block of plastic scintillator with dimensions \qtyproduct[product-units=power]{14x 8 x 8}{\milli\metre}. The scintillator used is Epic Crystal's EPS100, cut to size and wrapped with a layer of enhanced specular reflector (ESR) film and copper tape. The scintillators have a light yield of 65\% of that of anthracene at room temperature and emission peaks at a wavelength of \qty{423}{nm} \cite{EpicCrystalPlasticScint}. The modules are mounted next to each other on a copper plate with a gap of \qty{18}{mm} between the scintillators. A copper clamping plate is used to position the scintillators on top of the SiPMs. Figure~\ref{fig:reallifepictures} shows the wrapped plastic scintillators (left), SiPMs on PCBs mounted on the bracket (center) and the assembled muon detector (right). 

Figure~\ref{fig:full_schematic} shows the experimental setup for measuring the muon rate. The muon detector was mounted to the mixing chamber of the \emph{Bluefors} and later to that of \emph{Leiden-mK} to measure the muon rate at millikelvin temperatures. To confirm the successful operation of the detector, the measured rate was compared to that at room temperature and during heating/cooling. This was done because the muon flux should be independent of the detector temperature. 

\subsection{Muon rate data acquisition}\label{sec:muon_daq}
To measure the muon rate, the SiPMs were biased by the SMU within the operating voltage range (see \autoref{sec:VBD_VUL_vs_T}). The SiPM anodes were grounded, and the cathodes were connected to a readout box that stabilized the bias voltage and AC-coupled the output signals to an oscilloscope (see \autoref{fig:readoutboxcircuit}). The oscilloscope was triggered by a logical AND of the two SiPM signals, so that only coincident events were acquired. The waveform was saved at each trigger and processed in post to reject signals not caused by particles, such as pure noise. The resulting muon rates (see \autoref{sec:muon_rate}) are compared with expected rates from an analytical calculation based on the detector geometry and from a Geant4 Monte Carlo simulation (see Appendix \ref{sec:Simulation}).

To confirm that the detector correctly identifies muon events, we measured coincidences between our muon detector inside the dilution refrigerator and a commercially available educational muon detector made by mDetect~\cite{mdetect_muon}. This 2-layer detector was placed outside the cryostat, approximately below the internal detector, as shown in \autoref{fig:full_schematic}. An event was recorded only when both channels of the internal detector and both channels of the external detector triggered within the coincidence window.

\section{Results \label{sec:Results}}
\begin{figure}
    \centering
    \includegraphics[width=1\linewidth]{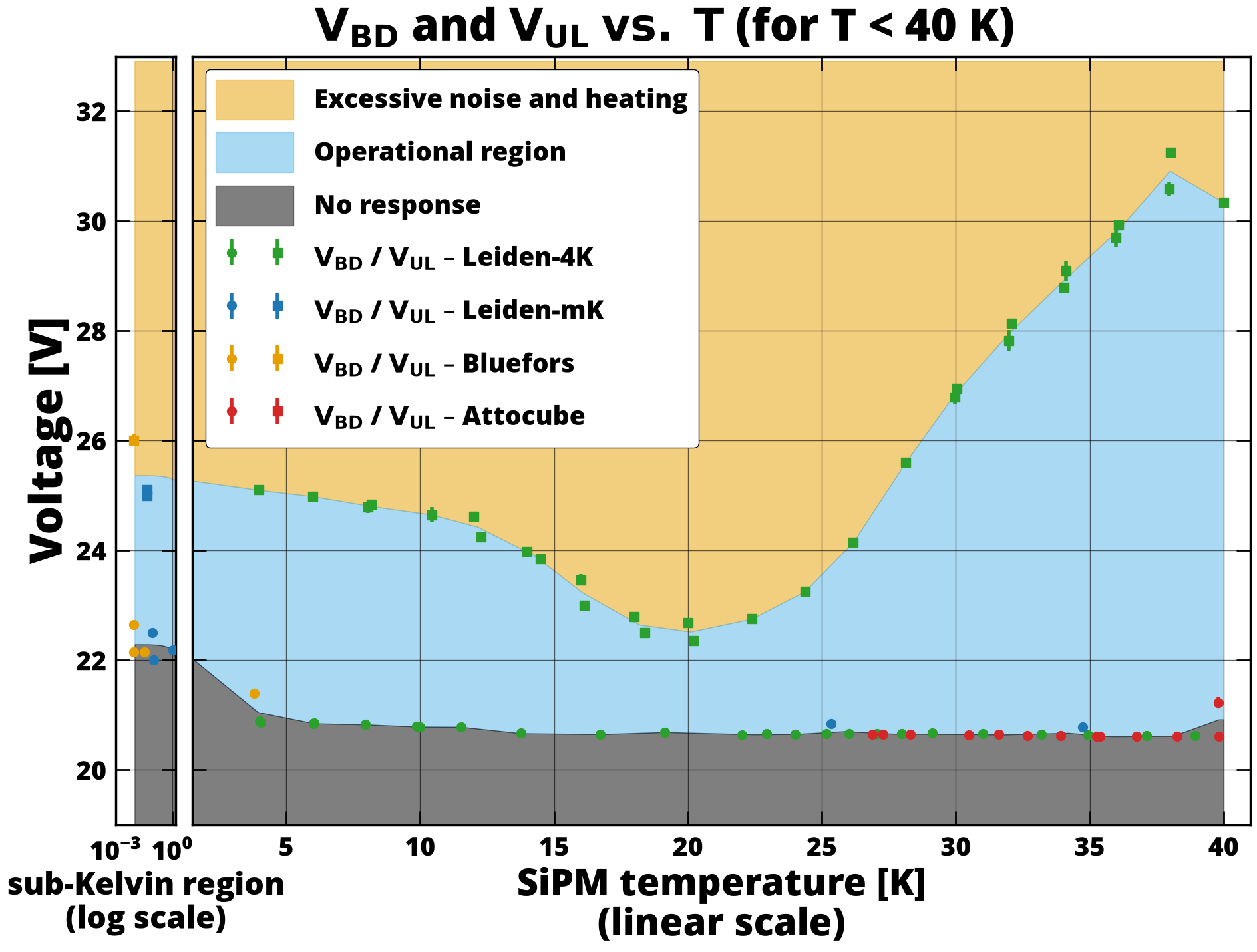}
    \caption{The operational region below \qty{40}{\kelvin}. Data points from I–V curves using different SiPMs from different measurement runs across all cryostats are included in the plot. The four cryostats are shown in the legend. The x-axis is divided into a sub-Kelvin temperature region to show the voltage threshold trend there distinctly. The shading in the plot is purely illustrative and was obtained by grouping the voltage data into temperature bins and performing a linear interpolation on the grouped data. The error bars are the error in the maximum of the fit to the $\mathrm{d}(\ln I)/\mathrm{d}V$ for $V_\mathrm{BD}$ and the voltage step size for $V_{\rm UL}$.} 
    \label{fig:VBD_VUL_vs_T}
\end{figure}    

\begin{figure}
    \centering
    \includegraphics[width=1\linewidth]{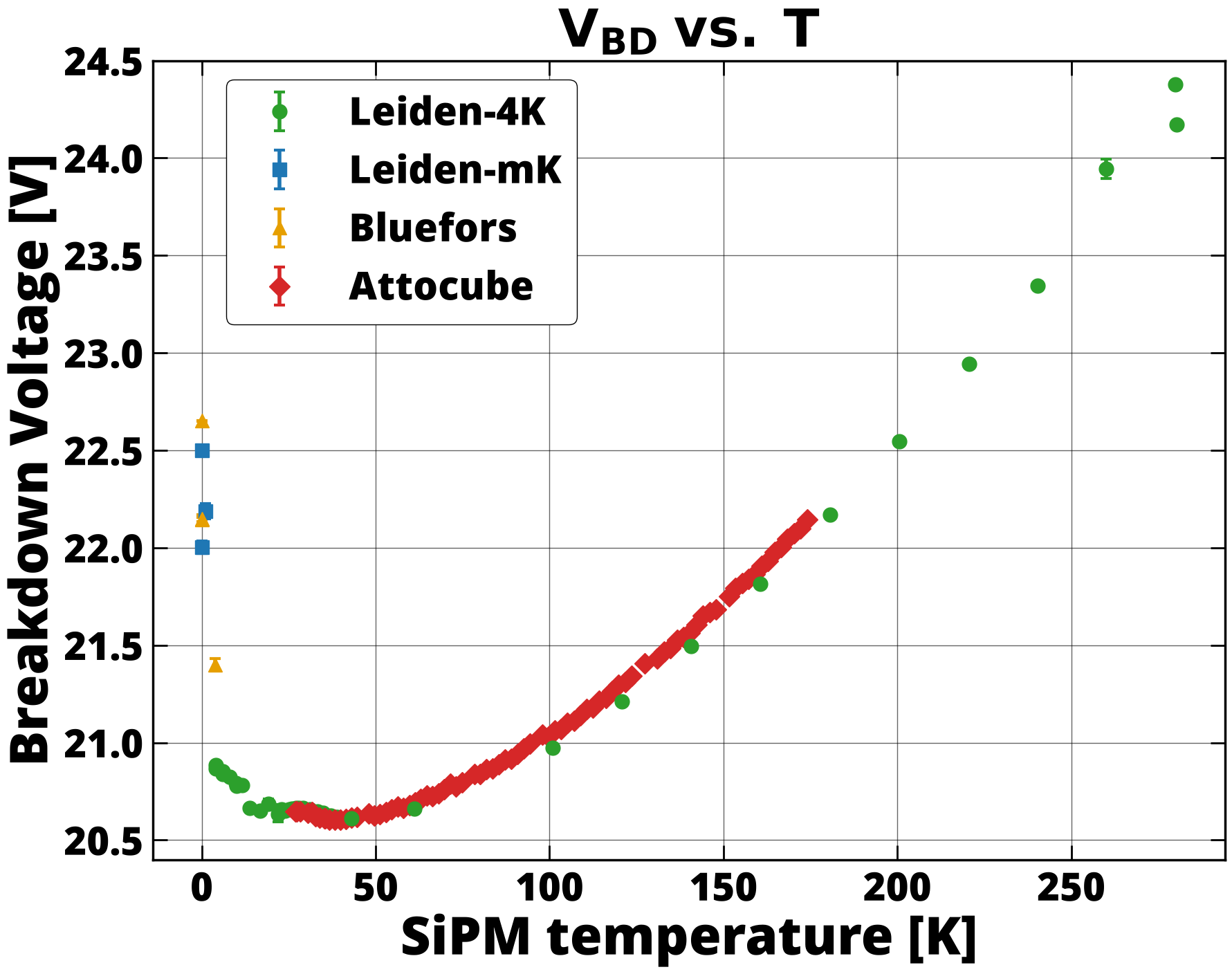}
    \caption{Breakdown voltage as a function of temperature, from millikelvin to room temperature. Each point is extracted from an I–V curve using a Langau fit. Different colors correspond to different cryostats; different SiPMs of the same model were used in each cryostat. The \emph{Attocube} data is the most accurate since there was a longer dwell time at each temperature and the temperature probe was directly coupled to the ground plane of the SiPM PCB. However, because of time constraints, the \emph{Attocube} measurements were not extended to room temperature, and they do not go below approximately \qty{25}{\kelvin} (the lowest temperature the cryostat reached with the SiPM mounted). Data from the other cryostats are included to cover the rest of the temperature range. The data below \qty{40}{\kelvin} are also used for the $V_{\mathrm{BD}}$ series in \autoref{fig:VBD_VUL_vs_T}.}
    \label{fig:VBD_vs_T}
\end{figure}

\begin{figure*}
    \centering
    \includegraphics[height=0.4\textheight]{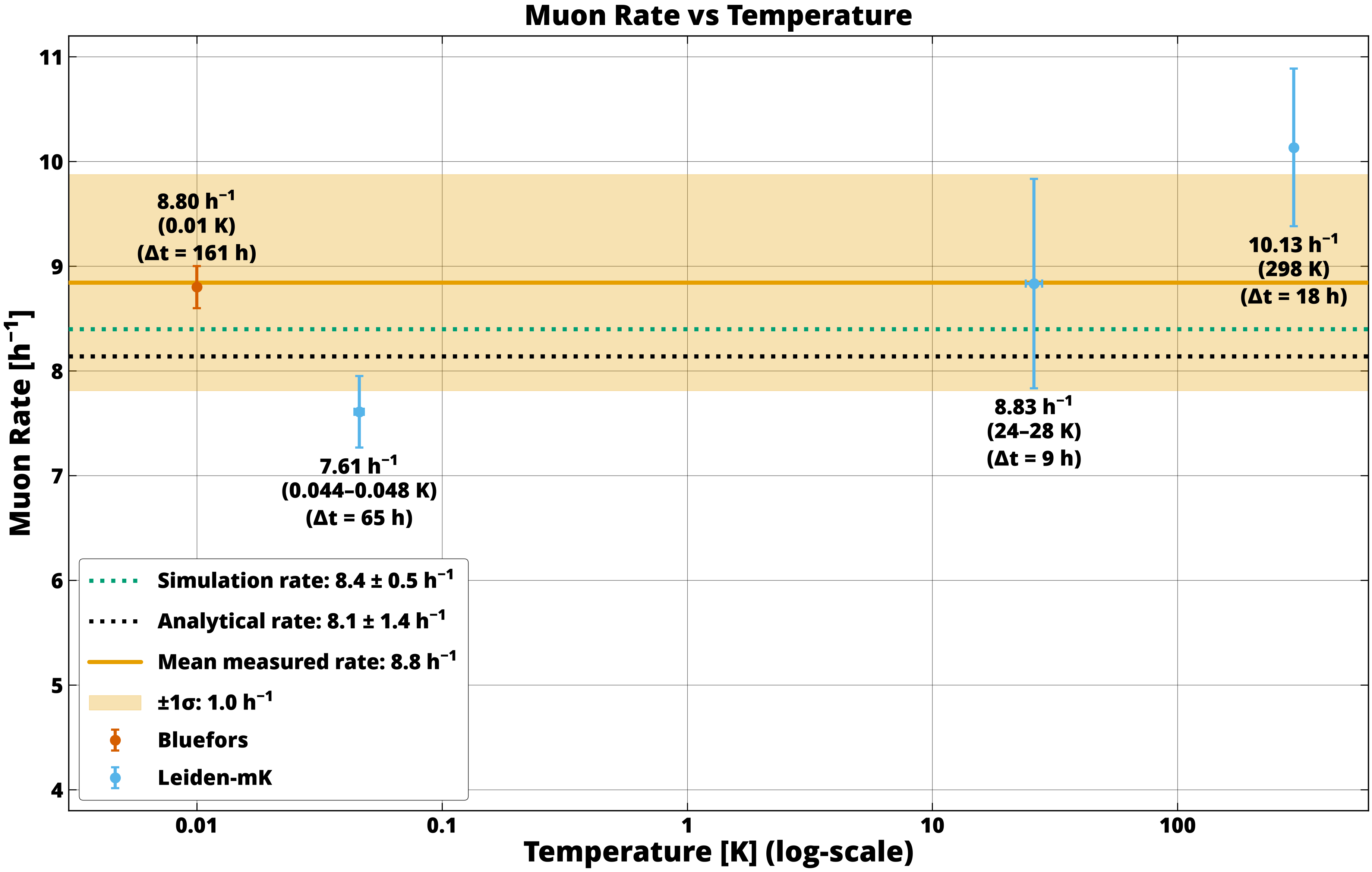}
    \caption{Muon rates measured at various temperatures inside dilution refrigerators (shown by the blue and orange points) with analytical (black dotted line) and simulated (green dotted line) rates. The temperatures over which the measurements in the dilution refrigerator were taken is highlighted under each point as well as the data acquisition time. The mean measured rate (orange solid line) and its $\pm 1\sigma$ bounds are also shown. The analytical rate uncertainty was calculated based on translational uncertainty of 1 mm along the vertical for both plastic scintillators.}
    \label{rates}
\end{figure*}

\subsection{SiPM characterization: $V_{\rm BD}(T)$ and $V_{\rm UL}(T)$\label{sec:VBD_VUL_vs_T}}
Figure~\ref{fig:VBD_VUL_vs_T} shows $V_{\mathrm{BD}}$ and $V_{\mathrm{UL}}$ from \qty{40}{\kelvin} down to \qty{9}{\milli\kelvin}, and \autoref{fig:VBD_vs_T} shows $V_{\mathrm{BD}}$ over the full range, from room temperature down to a few mK.

$V_{\mathrm{BD}}$ decreases from room temperature down to approximately \qty{40}{\kelvin} and increases again at lower temperatures. The same behavior has been reported for other SiPMs of different sizes and models~\cite{Hanski2025, gaoPathwaySubmeVDetection2026}.

The decrease in $V_{\mathrm{BD}}$ is caused by the reduction in phonon population as the sensor cools. Carriers undergo fewer lattice-scattering events and lose less energy while traveling through the depletion region. As a consequence, they can reach the energy required for impact ionization, and hence initiate an avalanche, at a lower electric field, corresponding to a lower $V_{\mathrm{BD}}$~\cite{Gutierrez2001}.

Below approximately \qty{40}{\kelvin}, however, dopant freeze-out becomes important. Electrons and holes increasingly remain bound to donor and acceptor states, respectively, reducing the fraction of ionized dopants. In the depletion region, this reduces the fixed space charge that generates the junction electric field. Consequently, a larger reverse bias is required to reach the electric field necessary for impact ionization, providing a possible explanation for the increase in $V_{\mathrm{BD}}$ at the lowest temperatures. Field-assisted ionization, such as the Poole--Frenkel effect, may partially counteract freeze-out by facilitating the release of carriers from impurity states~\cite{Gutierrez2001}.

\citet{Hanski2025} noted a correlation between the low-temperature behavior of $V_{\mathrm{BD}}$ and the temperature dependence of the electron mobility~\cite[Fig.~1.18]{Gutierrez2001}. However, the mobility shown there is the low-field mobility, whereas avalanche multiplication occurs in the high-field region of the junction. The similarity of the two temperature dependencies therefore does not necessarily imply that the change in mobility directly causes the change in $V_{\mathrm{BD}}$; both may instead be influenced by dopant freeze-out through different mechanisms.

The behavior of $V_{\mathrm{UL}}$ is more strongly non-monotonic, with a minimum around \qty{20}{\kelvin}. Similar behavior has been reported by  \citet{Zhang2022}, \citet{Cardini2014}, and \citet{Hanski2025}. \citet{Zhang2022} attributed the reduced operating range between approximately 20 and \qty{40}{\kelvin} to an increased probability of afterpulsing, which can develop into self-sustaining trains of avalanches. At lower temperatures, they observed the operating range to increase again and suggested that the longer trapping times of charge carriers delay the afterpulses sufficiently to interrupt this process.

Which processes drive this behavior at low temperature is not well understood. \citet{Hanski2025} proposed band-to-band or trap-assisted tunneling as a possible origin of the high-current transition. Band-to-band tunneling is only weakly temperature dependent compared with thermal generation of carriers, which is strongly suppressed at low temperature~\cite{Gutierrez2001}, and may therefore provide seed carriers that initiate avalanches at low temperatures. However, this alone does not explain the strongly non-monotonic temperature dependence of $V_{\mathrm{UL}}$. Other temperature-dependent processes, including trap-assisted carrier release, afterpulsing and crosstalk, may contribute to sustaining the avalanche activity. The present measurements do not distinguish between these mechanisms, and we therefore treat $V_{\mathrm{UL}}$ as an operational upper voltage limit, rather than naming it after a specific microscopic mechanism as done in~\citet{Hanski2025}.

\subsection{Muon coincidence rate}\label{sec:muon_rate}
If the detector works correctly, the measured muon rate should not depend on its temperature. We measured the muon rate at several temperatures, down to the mK range, and compared it with the rates obtained from an analytical calculation and from a Geant4 Monte Carlo simulation (see \autoref{rates}). The calculated and simulated rates include the attenuation by the building (26\%) since the dilution refrigerators are on the ground floor of a four-story building (see Appendix \ref{sec:Simulation}).

The mean measured rate is consistent with the analytical and simulated rates. The measured rate at each temperature stays within 15\% of the mean measured rate. As shown in \autoref{rates}, these rates were measured in different cryostats, using the same detector but slightly different experimental setups (e.g.\ different cables, PCBs and SiPMs, and scintillator positions differing by up to \qty{2}{\milli\metre}). These systematic errors account for the spread in the measured rates. 

\section{Double pulses below 8\ K}\label{sec:Double_pulses}
\begin{figure}    
\centering                  
\includegraphics[width=1\linewidth]{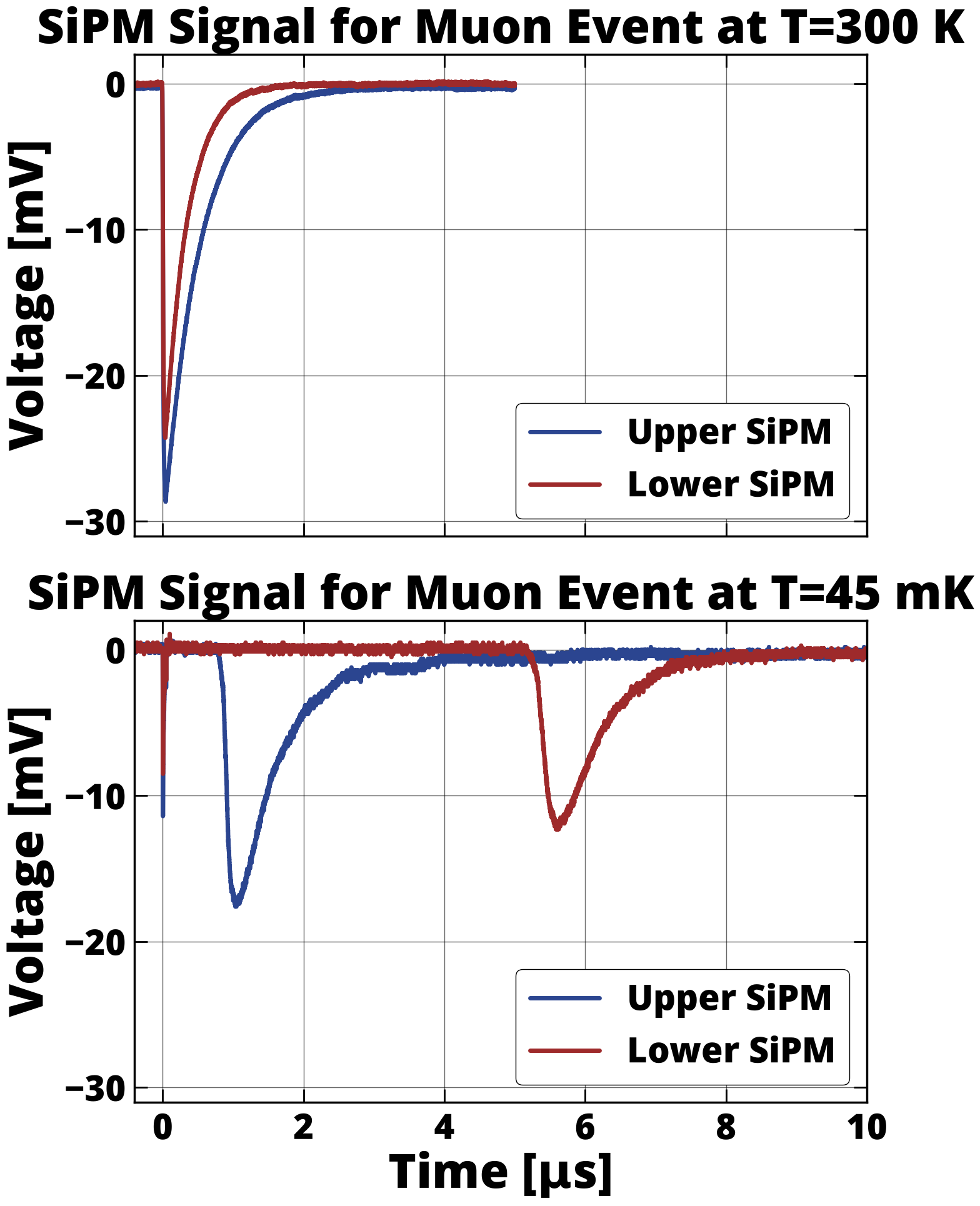}    
    \caption{The difference between normal SiPM signals above \qty{8}{\kelvin} (upper) and the double pulsing below \qty{8}{\kelvin} (lower), in response to muon events found using coincidence triggering. In the case of double pulsing, the narrow components of both SiPM output signals are aligned at \qty{0}{\micro\second} meaning the detector can still be used to tag muons with the same methodology. The room temperature traces were captured with a shorter record length, no additional secondary pulses were observed.}
    \label{waveform}
\end{figure}

When cooled below \qty{8}{\kelvin}, the SiPM output signal had a significantly different pulse shape. What is ordinarily a single pulse, with a fast rise time and a slower exponential decay, splits into two pulses. We refer to this as ``double pulsing'', and after significant investigation, we attribute the effect to the presence of small amounts of a superconducting material in the SiPM itself~\cite{swift2025superinductor}. Figure~\ref{waveform} shows examples of SiPM signals above and below \qty{8}{\kelvin}. When double pulsing occurs, we refer to the prompt, sharper pulse as the ``narrow component'' and the delayed, smoother pulse as the ``broad component''. 152 quadruple coincidence events tagged with two external muon detectors (see \autoref{sec:muon_daq}) confirm that even in the presence of double pulsing, the detector can detect muons. This is because the narrow component always appears promptly, at the arrival time of the scintillation light, so the coincidence timing is preserved. However, it is not yet known whether amplitude information (i.e. photon counting) can be extracted from double pulsing waveforms, since it is unclear which component carries this information.

We first excluded instrumental causes. The effect occurs regardless of the light source (LED or scintillation). It persists when using different instrumentation and readout electronics, and regardless of the type, temperature and length of the coaxial cables inside the cryostat. It occurs in all the cryostats except the \emph{Attocube}, which does not reach temperatures below \qty{8}{\kelvin} with a sample mounted.

\begin{figure}[t]    
\centering         
    \includegraphics[width=\linewidth]{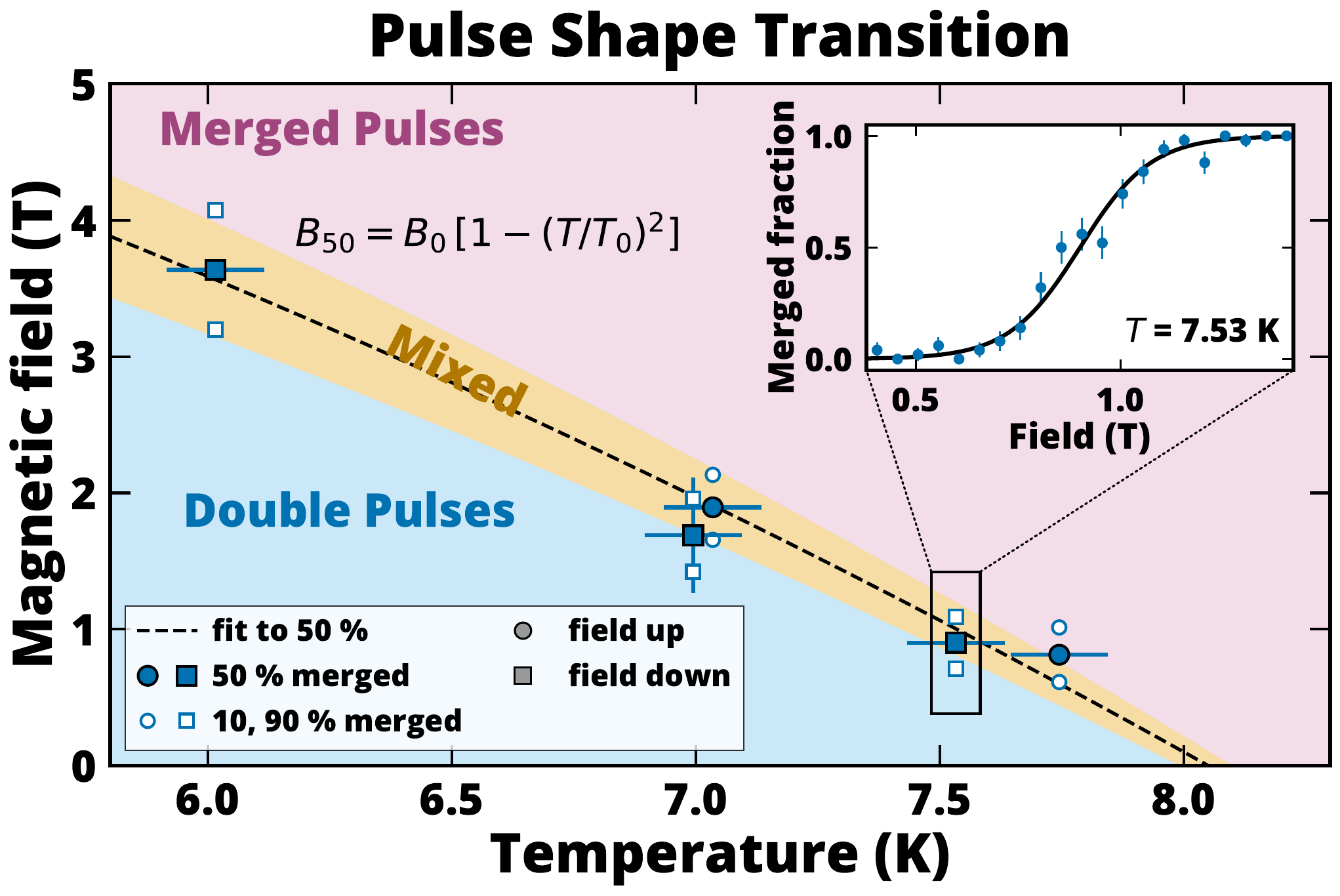}    
    \caption{Magnetic field--temperature phase diagram of the transition from double to merged pulses for one SiPM in \emph{Bluefors}. We ramp the magnetic field up and down to find the critical field values at which \qty{50}{\percent} ({\color[HTML]{0072B2}$\bullet$\,$\blacksquare$}, filled) and \qty{10}{\percent} and \qty{90}{\percent} ({\color[HTML]{0072B2}$\circ$\,$\square$}, open) of the pulses are merged; circles are for the field ramped up and squares for the field ramped down. The black dashed line is the parabolic fit to the \qty{50}{\percent} points. The same fits to the \qty{10}{\percent} and \qty{90}{\percent} points divide the diagram into the {\color[HTML]{0072B2}\textbf{double-pulse}} (blue), {\color[HTML]{B07800}\textbf{mixed}} (orange), and {\color[HTML]{A0457E}\textbf{merged-pulse}} (pink) regions. Inset: proportion of merged pulses as a function of magnetic for the ramp at $T=\qty{7.53}{\kelvin}$ (black box), with a logistic fit.}    \label{fig:pulse_shape_transition}
\end{figure}

To test whether a superconducting material is responsible, SiPMs were mounted on the cold finger in \emph{Bluefors} within the bore of a superconducting solenoid. At several fixed temperatures, the out-of-plane magnetic field was ramped up to \qty{4.5}{\tesla} and back down while the SiPM output signals and photocurrent were recorded. Above a temperature-dependent critical field, the narrow and broad components merge into a single pulse. At each B-field step we measured the fraction of merged pulses, which increased from 0 to 1 over a narrow B-field range (inset of \autoref{fig:pulse_shape_transition}). The B-fields at which \qty{10}{\percent}, \qty{50}{\percent} and \qty{90}{\percent} of the pulses are merged, plotted against temperature in \autoref{fig:pulse_shape_transition}, divide the diagram into three regions: double pulses below the \qty{10}{\percent} line, merged pulses above the \qty{90}{\percent} line, and a mixed region in between. All three lines are well described by the empirical critical-field law $B_0[1-(T/T_0)^2]$. The fit to the \qty{50}{\percent} points gives $T_c = \qty{8.05 \pm 0.08}{\kelvin}$ and $B_{c2}(0) = \qty{8.1 \pm 0.5}{\tesla}$ (see Appendix~\ref{sec:doublepulseappendix}). No significant difference was observed between ramping the magnetic field up or down.

Additionally, the forward voltage drop of the SiPMs was measured while varying the temperature of the cryostat around \qty{8}{K}. A sharp increase in the voltage drop was observed as the temperature increased above \qty{8}{K}, indicating a sudden increase in the impedance of the device (see Appendix~\ref{sec:doublepulseappendix}).

The superconducting material has not been identified with certainty. Thin films such as \ce{NbTi}, \ce{Nb} and \ce{TiN} are used in semiconductor fabrication as adhesion layers and diffusion barriers~\cite{wittmer_properties_1985,sherman_growth_1990}, and are type-II superconductors with critical temperatures compatible with the observed transition.
\FloatBarrier

\section{Conclusions \label{sec:Conclusions}}
We have built and characterized a low-cost, compact muon detector for operation inside a dilution refrigerator, consisting of two EPS100 plastic scintillator, each read out by an onsemi MicroFC-60035 SiPM, operated in coincidence. The SiPMs, scintillators and PCBs together cost less than 100~USD and this paper provides all the information needed to reproduce it, including the detector geometry, the readout circuitry and the operating bias range at each temperature. We have also brought together, for the first time, all available studies of SiPMs, scintillator--SiPM detectors and KIDs at cryogenic temperatures, showing which detectors have been operated, at what temperatures, and what was measured (see \autoref{fig:LITREVIEWPLOT}).

The SiPMs were characterized from room temperature down to \qty{9}{\milli\kelvin}. The breakdown voltage $V_{\mathrm{BD}}$ decreases from room temperature down to approximately \qty{40}{\kelvin} and increases again at lower temperatures, while the upper-limit voltage $V_{\mathrm{UL}}$ is strongly non-monotonic, with a minimum around \qty{20}{\kelvin}. Together, the two voltages define the bias range available for operation at each temperature. Between \qty{46}{\kelvin} and room temperature, we also measured the voltage gain, the optical crosstalk probability, the SPE resolution, and the PDE relative to room temperature.

Operated as a muon counter in two dilution refrigerators, the detector measures a coincidence rate that does not change with temperature from \qty{300}{\kelvin} to \qty{10}{\milli\kelvin} and agrees with analytical and Monte Carlo predictions. 

Below approximately \qty{8}{\kelvin}, each muon event produces two pulses, a narrow and a broad component. Measurements as a function of temperature and magnetic field up to \qty{4.5}{\tesla} show that the transition from double to merged pulses follows the phase diagram of a type-II superconductor, linking the effect to a superconducting material in the SiPM, possibly \ce{NbTi, TiN, or $^*$Nb}. Since the narrow component always appears at the arrival time of the scintillation light, the coincidence timing is preserved, and quadruple coincidence measurements with external detectors confirm that muons are still counted reliably in this regime.

Future work will address the mechanism by which the superconducting state produces the second pulse, and whether photon counting information can be extracted from double pulse waveforms.

As this effect has not, to our knowledge, been reported before, we hope that this work will make it known to the wider physics community, and we welcome suggestions that could help explain it. 

We will test other SiPM models, starting with the onsemi J-series MicroFJ-60035-TSV, which has the same active area and microcell size as the C-series sensor used here and for which, to our knowledge, no double pulsing has been reported. We will also test other scintillating materials, such as \ce{CaWO4}, \ce{CdWO4} and \ce{CsI}. 

Finally, we plan to install the detector alongside a qubit array, with the scintillators placed closer together to increase the solid-angle coverage, to tag muon interactions and study how these tags can inform quantum error correction.

\section*{Acknowledgments}
This research was supported by the Australian Government through the Australian Research Council Centre of Excellence for Dark Matter Particle Physics (award number CE200100008) and the Australian Research Council Discovery Project DP260101028. Additional support was provided by the University of Sydney International Stipend Scholarship and the Physics Foundation Scholarship.

We thank David Reilly for access to his laboratory, dilution refrigerators and equipment, and Brendan Harlech-Jones for support with operating the dilution refrigerators. We also thank Boris Kuhlmey for access to his laboratory and \emph{Attocube}, Simon Fleming for directing us to the \emph{Attocube}, Nick Proschogo for help with repairing the helium line, and Adriano Di Giovanni for providing the ESR film. We are grateful to Joey Liang for help with repairing the \emph{Attocube}, for help with the literature review and for useful discussions; Jonathan Zhu for help with the \emph{Attocube}, the network and Raspberry Pi setup, and soldering of the SiPMs; Artem Knyazev for his contribution to the early stages of the project; and Yousef Abdelrauf for automating the $I$--$V$ measurements. The style of the plots is adapted from the GitHub repository of Ciaran O'Hare~\cite{AxionLimits}, whom we thank for setting new aesthetic standards. Finally, we thank the staff of the School of Physics workshop: Caleb Gudu, Anil Pattarath, Nagaraja (Raj) Panduranga and Poulose Arakkudiyil.

\section*{Author contributions}
Tengiz Ibrayev coordinated the measurements and the day-to-day running of the project, set up the data collection procedures for the \emph{Attocube} and the dilution refrigerators, and carried out a large part of the data analysis.

Danish Khan designed the PCBs and the readout box, set up and collected data for most of the dilution refrigerator runs, and contributed to the literature review and data analysis.

Alex Herscovics wrote the automation code for the $I$--$V$ characterization in the \emph{Attocube}, performed the analysis of the operating voltages, produced schematics and figures, carried out the mechanical design of the original mounting bracket, and helped set up most of the runs and collect the data.

Komal Kataruka helped set up, collect data, and analyze the gain and breakdown voltage measurements in the \emph{Attocube}.

Matilda Lawton led the Monte Carlo simulation, carried out the literature review, contributed to the analysis of the muon rate, performed the pulse area and time delay analysis, and produced figures.

Sandhya Sharma contributed to the literature review and to the analysis of the muon rate, produced figures and schematics, helped with data collection, and designed a mounting bracket for \textit{Bluefors} cold finger.

Steven Waddy operated and was responsible for the refrigerators, helped to assemble the experimental setup, helped design the mounting bracket, and proposed superconductivity as the origin of the double pulses.

Laura Manenti conceived and led the project, secured the funding, supervised and guided the team, defined the structure and narrative of the paper, and wrote most of the manuscript.

Tengiz Ibrayev, Danish Khan, Alex Herscovics, Matilda Lawton and Sandhya Sharma contributed to writing of the manuscript. All authors reviewed the manuscript.

\section*{Use of AI tools}
ChatGPT and Claude Code were used to correct grammar and typographical errors in the manuscript and to assist in searching the literature. 
They assisted in writing code for automating data collection, performing analysis on $I$--$V$ curves and finger plots, refining and stylizing plots. They also monitored the experimental runs and flagged possible malfunctions. 
All experimental work and interpretations were carried out by the authors; all data analysis was verified by the authors. Notably, none of the language models were able to find an explanation in the literature (or on their own) for the double pulses reported in this work.

\bibliographystyle{bibstyle}
\bibliography{main,appendix_sipm}

@article{Gallego2013,
  author  = {Gallego, L. and Rosado, J. and Blanco, F. and Arqueros, F.},
  title   = {Modeling crosstalk in silicon photomultipliers},
  journal = {JINST},
  volume  = {8},
  pages   = {P05010},
  year    = {2013},
  doi     = {10.1088/1748-0221/8/05/P05010},
  eprint  = {1302.1455},
  archivePrefix = {arXiv}
}

@article{Chmill2017,
  author  = {Chmill, V. and Garutti, E. and Klanner, R. and Nitschke, M. and Schwandt, J.},
  title   = {On the characterisation of {SiPMs} from pulse-height spectra},
  journal = {Nucl. Instrum. Meth. A},
  volume  = {854},
  pages   = {70--81},
  year    = {2017},
  doi     = {10.1016/j.nima.2017.02.049},
  eprint  = {1609.01181},
  archivePrefix = {arXiv}
}

@article{Vinogradov2012,
  author  = {Vinogradov, S.},
  title   = {Analytical models of probability distribution and excess noise factor of solid state photomultiplier signals with crosstalk},
  journal = {Nucl. Instrum. Meth. A},
  volume  = {695},
  pages   = {247--251},
  year    = {2012},
  doi     = {10.1016/j.nima.2011.11.086},
  eprint  = {1109.2014},
  archivePrefix = {arXiv}
}

@article{Rehbein2023,
  author  = {Rehbein, F. and Bretz, T. and Biland, A. and Alfaro, R. and Audehm, J. and Do, G. and Gonz{\'a}lez, M. M. and P{\'e}rez Araujo, Y. F. and Schaufel, M. and Serna-Franco, J. and Torres, I.},
  title   = {Calibration of 122 {SensL} {MicroFJ-60035} {SiPMs} and the Reduction of Optical Cross-Talk Due to Coupled Light Guides},
  journal = {Rev. Mex. Astron. Astrofis.},
  volume  = {59},
  number  = {2},
  pages   = {379--388},
  year    = {2023},
  doi     = {10.22201/ia.01851101p.2023.59.02.14}
}

@article{Kjaergaard2020,
  author  = {Kjaergaard, Morten and Schwartz, Mollie E. and Braum{\"u}ller, Jochen
             and Krantz, Philip and Wang, Joel I.-J. and Gustavsson, Simon
             and Oliver, William D.},
  title   = {Superconducting Qubits: Current State of Play},
  journal = {Annual Review of Condensed Matter Physics},
  volume  = {11},
  pages   = {369--395},
  year    = {2020},
  doi     = {10.1146/annurev-conmatphys-031119-050605}
}

@software{AxionLimits,
  author    = {O'Hare, Ciaran},
  title     = {cajohare/AxionLimits: AxionLimits},
  year      = {2020},
  publisher = {Zenodo},
  version   = {v1.0},
  doi       = {10.5281/zenodo.3932430},
  url       = {https://cajohare.github.io/AxionLimits/}
}

@article{Vepsalainen2020,
  author  = {Veps{\"a}l{\"a}inen, Antti P. and Karamlou, Amir H. and Orrell, John L.
             and Dogra, Akshunna S. and Loer, Ben and Vasconcelos, Francisca
             and Kim, David K. and Melville, Alexander J. and Niedzielski, Bethany M.
             and Yoder, Jonilyn L. and Gustavsson, Simon and Formaggio, Joseph A.
             and VanDevender, Brent A. and Oliver, William D.},
  title   = {Impact of ionizing radiation on superconducting qubit coherence},
  journal = {Nature},
  volume  = {584},
  pages   = {551--556},
  year    = {2020},
  doi     = {10.1038/s41586-020-2619-8}
}

@article{Wilen2021,
  author  = {Wilen, C. D. and Abdullah, S. and Kurinsky, N. A. and Stanford, C.
             and Cardani, L. and D'Imperio, G. and Tomei, C. and Faoro, L.
             and Ioffe, L. B. and Liu, C. H. and Opremcak, A. and Christensen, B. G.
             and DuBois, J. L. and McDermott, R.},
  title   = {Correlated charge noise and relaxation errors in superconducting qubits},
  journal = {Nature},
  volume  = {594},
  pages   = {369--373},
  year    = {2021},
  doi     = {10.1038/s41586-021-03557-5}
}

@article{McEwen2022,
  author  = {McEwen, Matt and Faoro, Lara and Arya, Kunal and Dunsworth, Andrew
             and Huang, Trent and Kim, Seon and Burkett, Brian and Fowler, Austin
             and Arute, Frank and others},
  title   = {Resolving catastrophic error bursts from cosmic rays in large arrays
             of superconducting qubits},
  journal = {Nature Physics},
  volume  = {18},
  pages   = {107--111},
  year    = {2022},
  doi     = {10.1038/s41567-021-01432-8}
}

@article{Li2025,
  author  = {Li, Xuegang and Wang, Junhua and Jiang, Yao-Yao and Xue, Guang-Ming
             and Cai, Xiaoxia and Zhou, Jun and Gong, Ming and Liu, Zhao-Feng
             and Zheng, Shuang-Yu and Ma, Deng-Ke and Chen, Mo and Sun, Wei-Jie
             and Yang, Shuang and Yan, Fei and Jin, Yi-Rong and Zhao, S. P.
             and Ding, Xue-Feng and Yu, Hai-Feng},
  title   = {Cosmic-ray-induced correlated errors in superconducting qubit array},
  journal = {Nature Communications},
  volume  = {16},
  number  = {1},
  pages   = {4677},
  year    = {2025},
  doi     = {10.1038/s41467-025-59778-z}
}

@misc{mdetect_muon,
  author       = {{mDetect}},
  title        = {mDetect -- Seeing through the unseeable},
  howpublished = {\url{https://mdetect.com.au/}},
  note         = {Accessed: 2026-10-13}
}

@article{Karatsu2019,
  author  = {Karatsu, K. and Endo, A. and Bueno, J. and de Visser, P. J.
             and Barends, R. and Thoen, D. J. and Murugesan, V. and Tomei, N.
             and Baselmans, J. J. A.},
  title   = {Mitigation of cosmic ray effect on microwave kinetic inductance
             detector arrays},
  journal = {Applied Physics Letters},
  volume  = {114},
  number  = {3},
  pages   = {032601},
  year    = {2019},
  doi     = {10.1063/1.5052419}
}

@article{Mariani2026,
  author    = {Mariani, Ambra and Cardani, Laura and Bal, Mustafa and Casali, Nicola
               and Colantoni, Ivan and Cruciani, Angelo and Del Castello, Giorgio
               and Delicato, Daniele and De Dominicis, Francesco
               and del Gallo Raccagiovine, Matteo and Folcarelli, Matteo
               and Garattoni, Sabrina and Grassellino, Anna
               and Khan Yasir Raja, Mehmood and Pettinacci, Valerio
               and Ressa, Alberto and Roy, Tanay and Vignati, Marco
               and van Zanten, David},
  title     = {A cryogenic muon tagging system based on kinetic inductance detectors
               for superconducting quantum processors},
  journal   = {Quantum Science and Technology},
  volume    = {11},
  number    = {2},
  pages     = {025053},
  year      = {2026},
  publisher = {IOP Publishing},
  doi       = {10.1088/2058-9565/ae5fcc}
}

@article{Erhart2024,
  author  = {Erhart, A. and Wagner, V. and Wex, A. and Goupy, C. and Lhuillier, D.
             and Namuth, E. and Nones, C. and Rogly, R. and Savu, V. and Schwarz, M.
             and Strauss, R. and Vivier, M. and Abele, H. and Angloher, G.
             and Bento, A. and Burkhart, J. and Canonica, L. and Cappella, F.
             and Casali, N. and Cerulli, R. and Cruciani, A. and del Castello, G.
             and del Gallo Roccagiovine, M. and Doblhammer, A. and Dorer, S.
             and Friedl, M. and Garai, A. and Ghete, V. M. and Hauff, D.
             and Jeanneau, F. and Jericha, E. and Kaznacheeva, M. and Kinast, A.
             and Kluck, H. and Langenk{\"a}mper, A. and Lasserre, T. and Mancuso, M.
             and Martin, R. and Mauri, B. and Mazzolari, A. and Mazzucato, E.
             and Neyrial, H. and Oberauer, L. and Ortmann, T. and Pattavina, L.
             and Peters, L. and Petricca, F. and Potzel, W. and Pr{\"o}bst, F.
             and Pucci, F. and Reindl, F. and Romagnoni, M. and Rothe, J.
             and Schermer, N. and Schieck, J. and Sch{\"o}nert, S.
             and Schwertner, C. and Scola, L. and Soum-Sidikov, G.
             and Stodolsky, L. and Tamisari, M. and Tomei, C. and Vignati, M.},
  title   = {A plastic scintillation muon veto for sub-Kelvin temperatures},
  journal = {The European Physical Journal C},
  volume  = {84},
  number  = {1},
  pages   = {70},
  year    = {2024},
  doi     = {10.1140/epjc/s10052-023-12375-0}
}

@article{Kemp2026,
  author    = {Kemp, A. and Autti, S. and Bloomfield, E. and Casey, A.
               and Darvishi, N. and Doling, D. and Eng, N. and Franchini, P.
               and Haley, R. P. and Heikkinen, P. J. and Jennings, A.
               and Koulosousas, S. and Leason, E. and Levitin, L. V.
               and March-Russell, J. and Mayer, A. and Monroe, J.
               and M{\"u}nstermann, D. and Noble, M. T. and Prance, J. R.
               and Rojas, X. and Salmon, T. and Saunders, J. and Smirnov, J.
               and Smith, R. and Thompson, M. D. and Thomson, A. and Ting, A.
               and Tsepelin, V. and West, S. M. and Whitehead, L. and Zmeev, D. E.
               and {The QUEST-DMC Collaboration}},
  title     = {Characterisation of silicon photomultipliers in a dilution
               refrigerator down to 9.4~mK towards a cryogenic cosmic-ray muon
               veto system},
  journal   = {Journal of Instrumentation},
  volume    = {21},
  number    = {05},
  pages     = {P05008},
  year      = {2026},
  publisher = {IOP Publishing},
  doi       = {10.1088/1748-0221/21/05/P05008}
}

@article{Hanski2025,
  author  = {Hanski, Otto and Kiilerich, Tom and Ahopelto, Sampsa
             and Semakin, Aleksei and Ahokas, Janne and Dvornichenko, Viacheslav
             and Vasiliev, Sergey},
  title   = {Performance of silicon photomultipliers from room temperature to
             below 200~mK},
  journal = {The European Physical Journal D},
  volume  = {79},
  number  = {7},
  pages   = {89},
  year    = {2025},
  doi     = {10.1140/epjd/s10053-025-01034-6}
}

@article{Zhang2022,
  author    = {Zhang, J. and Goeldi, D. and Iwai, R. and Sakurai, M.
               and S{\'o}t{\'e}r, A.},
  title     = {Scintillation detectors with silicon photomultiplier readout in a
               dilution refrigerator at temperatures down to 0.2~K},
  journal   = {Journal of Instrumentation},
  volume    = {17},
  number    = {06},
  pages     = {P06024},
  year      = {2022},
  publisher = {IOP Publishing},
  doi       = {10.1088/1748-0221/17/06/P06024}
}

@article{Collazuol2011,
  author  = {Collazuol, G. and Bisogni, M. G. and Marcatili, S. and Piemonte, C.
             and Del Guerra, A.},
  title   = {Studies of silicon photomultipliers at cryogenic temperatures},
  journal = {Nuclear Instruments and Methods in Physics Research Section A},
  volume  = {628},
  number  = {1},
  pages   = {389--392},
  year    = {2011},
  doi     = {10.1016/j.nima.2010.07.008}
}

@article{Acerbi2017,
  author  = {Acerbi, Fabio and Davini, Stefano and Ferri, Alessandro
             and Galbiati, Cristiano and Giovanetti, Graham and Gola, Alberto
             and Korga, George and Mandarano, Andrea and Marcante, Marco
             and Paternoster, Giovanni and Piemonte, Claudio and Razeto, Alessandro
             and Regazzoni, Veronica and Sablone, Davide and Savarese, Claudio
             and Zappal{\'a}, Gaetano and Zorzi, Nicola},
  title   = {Cryogenic Characterization of {FBK} {HD} Near-{UV} Sensitive {SiPMs}},
  journal = {IEEE Transactions on Electron Devices},
  volume  = {64},
  number  = {2},
  pages   = {521--526},
  year    = {2017},
  doi     = {10.1109/TED.2016.2641586}
}

@article{Aalseth2017,
  author    = {Aalseth, C. E. and Acerbi, F. and Agnes, P. and Albuquerque, I. F. M.
               and Alexander, T. and others},
  title     = {Cryogenic Characterization of {FBK} {RGB}-{HD} {SiPMs}},
  journal   = {Journal of Instrumentation},
  volume    = {12},
  number    = {09},
  pages     = {P09030},
  year      = {2017},
  publisher = {IOP Publishing},
  doi       = {10.1088/1748-0221/12/09/P09030}
}

@article{Lightfoot2008,
  author    = {Lightfoot, P. K. and Barker, G. J. and Mavrokoridis, K.
               and Ramachers, Y. A. and Spooner, N. J. C.},
  title     = {Characterisation of a silicon photomultiplier device for applications
               in liquid argon based neutrino physics and dark matter searches},
  journal   = {Journal of Instrumentation},
  volume    = {3},
  number    = {10},
  pages     = {P10001},
  year      = {2008},
  publisher = {IOP Publishing},
  doi       = {10.1088/1748-0221/3/10/P10001}
}

@article{Klanner2019,
  author  = {Klanner, Robert},
  title   = {Characterisation of {SiPMs}},
  journal = {Nuclear Instruments and Methods in Physics Research Section A},
  volume  = {926},
  pages   = {36--56},
  year    = {2019},
  doi     = {10.1016/j.nima.2018.11.083}
}

@techreport{OnsemiMicroC2022,
  author      = {{onsemi}},
  title       = {{MicroC}-Series Silicon Photomultipliers ({SiPM}) Data Sheet},
  institution = {Semiconductor Components Industries, LLC},
  year        = {2022},
  month       = feb,
  url         = {https://www.onsemi.com/pdf/datasheet/microc-series-d.pdf},
  note        = {Originally published 2014}
}

@techreport{EpicCrystalPlasticScint,
  author      = {{Epic Crystal Co., Ltd.}},
  title       = {{Plastic Scintillator} Data Sheet},
  institution = {Epic Crystal Co., Ltd.},
  year        = {2026},
  month       = feb,
  url         = {https://www.epic-crystal.com/data/upload/20260206/69855ae234086.pdf}
}

@article{wiesinger2023,
	title = {Trap-integrated fluorescence detection with silicon photomultipliers for sympathetic laser cooling in a cryogenic {Penning} trap},
	volume = {94},
	issn = {0034-6748},
	url = {https://doi.org/10.1063/5.0170629},
	doi = {10.1063/5.0170629},
	number = {12},
	urldate = {2026-05-08},
	journal = {Review of Scientific Instruments},
	author = {Wiesinger, M. and Stuhlmann, F. and Bohman, M. and Micke, P. and Will, C. and Yildiz, H. and Abbass, F. and Arndt, B. P. and Devlin, J. A. and Erlewein, S. and Fleck, M. and Jäger, J. I. and Latacz, B. M. and Schweitzer, D. and Umbrazunas, G. and Wursten, E. and Blaum, K. and Matsuda, Y. and Mooser, A. and Quint, W. and Soter, A. and Walz, J. and Smorra, C. and Ulmer, S.},
	month = dec,
	year = {2023},
	pages = {123202},
}

@ARTICLE{Cardini2014,
  author={Cardini, Alessandro and Fanti, Viviana and Lai, Adriano},
  journal={IEEE Transactions on Nuclear Science}, 
  title={Operation of Multi-Pixel Photon Counters Down to Liquid Helium Temperatures}, 
  year={2014},
  volume={61},
  number={5},
  pages={2672-2677},
  doi={10.1109/TNS.2014.2357254}}

@article{Mirzakhani2026,
  author  = {Mirzakhani, M. and Mahapatra, R. and Platt, M.},
  title   = {Characterization of a {CsI}({Tl}) scintillator coupled to a {SiPM}
             at cryogenic temperatures},
  journal = {Nuclear Instruments and Methods in Physics Research Section A},
  volume  = {1088},
  pages   = {171520},
  year    = {2026},
  doi     = {10.1016/j.nima.2026.171520}
}

@inbook{Gutierrez2001,
author = {Gutiérrez-D, E.A.},
year = {2001},
month = {12},
pages = {},
title = {Chapter 1. Physics of Silicon at Cryogenic Temperatures},
isbn = {9780123106759},
doi = {10.1016/B978-012310675-9/50002-5}
}

@article{Chiles2022,
  author  = {Chiles, Jeff and Charaev, Ilya and Lasenby, Robert
             and Baryakhtar, Masha and Huang, Junwu and Roshko, Alexana
             and Burton, George and Colangelo, Marco and Van Tilburg, Ken
             and Arvanitaki, Asimina and Nam, Sae Woo and Berggren, Karl K.},
  title   = {New Constraints on Dark Photon Dark Matter with Superconducting
             Nanowire Detectors in an Optical Haloscope},
  journal = {Physical Review Letters},
  volume  = {128},
  number  = {23},
  pages   = {231802},
  year    = {2022},
  doi     = {10.1103/PhysRevLett.128.231802}
}

@article{QROCODILE2025,
  author  = {Baudis, L. and Bismark, A. and Brugger, N. and Capelli, C.
             and Charaev, I. and Cuenca-Garc{\'i}a, J. J. and Hadas, G. D.
             and Hochberg, Y. and Hohmann, J. and Kavner, A. and Koos, C.
             and Kuzmin, A. and Lehmann, B. V. and N{\"a}geli, S.
             and Neupert, T. and Penning, B. and Ram{\'i}rez Garc{\'i}a, D.
             and Schilling, A.},
  title   = {First Sub-{MeV} Dark Matter Search with the {QROCODILE} Experiment
             Using Superconducting Nanowire Single-Photon Detectors},
  journal = {Physical Review Letters},
  volume  = {135},
  number  = {8},
  pages   = {081002},
  year    = {2025},
  doi     = {10.1103/4hb6-f6jl}
}

@article{Groom2001,
  author  = {Groom, D. E. and Mokhov, N. V. and Striganov, S. I.},
  title   = {Muon Stopping Power and Range Tables 10 {MeV}--100 {TeV}},
  journal = {Atomic Data and Nuclear Data Tables},
  volume  = {78},
  number  = {2},
  pages   = {183--356},
  year    = {2001},
  doi     = {10.1006/adnd.2001.0861}
}

@article{swift2025superinductor,
  title={A superinductor in a deep sub-micron integrated circuit},
  author={Swift, Thomas H and Olivieri, Fabio and Aizpurua-Iraola, Gorka and Kirkman, James and Noah, Grayson M and de Kruijf, Mathieu and von Horstig, Felix-Ekkehard and Gomez-Saiz, Alberto and Morton, John JL and Gonzalez-Zalba, M Fernando},
  journal={arXiv preprint arXiv:2507.13202},
  year={2025}
}

@misc{gaoPathwaySubmeVDetection2026,
  author        = {Gao, Aiqin and Wang, Hengyu and Li, Xuegang and Wang, Junhua and Lv, Junguang
                   and Qu, Guopu and Cao, Lei and Sun, Xilei and Guo, Yiming},
  title         = {Exploring the performance of {SiPM} at cryogenic temperature for the sub-{meV} threshold detector},
  year          = {2025},
  eprint        = {2509.11954v2},
  archivePrefix = {arXiv},
  primaryClass  = {physics.ins-det},
  doi           = {10.48550/arXiv.2509.11954},
  note          = {Version 2; the preprint was later withdrawn by the authors (v4, 2026)}
}

@article{darroch2025measuring,
  title={Measuring dark currents in multiple cryogenic SiPMs with sub-pA sensitivity using an automated IV multiplexer},
  author={Darroch, Lucas and Egan, Eamon and Leclerc, Marc-Antoine and McElroy, Thomas and Brunner, Thomas},
  journal={IEEE Transactions on Instrumentation and Measurement},
  year={2025},
  publisher={IEEE}
}

@article{eckertCharacterisationStudiesSilicon2010,
  title = {Characterisation Studies of Silicon Photomultipliers},
  author = {Eckert, Patrick and Schultz-Coulon, Hans-Christian and Shen, Wei and Stamen, Rainer and Tadday, Alexander},
  date = {2010-08-11},
  year = {2010},
  journal = {Nucl. Instrum. Meth. A},
  journaltitle = {Nuclear Instruments and Methods in Physics Research Section A: Accelerators, Spectrometers, Detectors and Associated Equipment},
  shortjournal = {Nuclear Instruments and Methods in Physics Research Section A: Accelerators, Spectrometers, Detectors and Associated Equipment},
  volume = {620},
  number = {2--3},
  pages = {217--226},
  issn = {0168-9002},
  doi = {10.1016/j.nima.2010.03.169},
  url = {https://www.sciencedirect.com/science/article/pii/S0168900210008156}
}

@online{WhatSiPMHowa,
  title = {What Is an {{SiPM}} and How Does It Work? | {{Hamamatsu Photonics}}},
  shorttitle = {What Is an {{SiPM}} and How Does It Work?},
  url = {https://hub.hamamatsu.com/us/en/technical-notes/mppc-sipms/what-is-an-SiPM-and-how-does-it-work.html},
  langid = {english}
}

@techreport{onsemi_and9770,
  author      = {{onsemi}},
  title       = {Introduction to the Silicon Photomultiplier (SiPM)},
  institution = {onsemi},
  number      = {AND9770/D},
  type        = {Application Note},
  url         = {https://www.onsemi.com/pub/collateral/and9770-d.pdf},
  date = {2023-08},
  year = {2023},
  month = aug,
  note = {{Rev.~9}}
}

@techreport{PDGCosmicRays2022,
  author      = {Beatty, J. J. and Matthews, J. and Wakely, S. P.},
  title       = {Cosmic Rays},
  institution = {Particle Data Group},
  year        = {2022},
  number      = {30},
  note        = {Revised October 2021},
  url         = {https://pdg.lbl.gov/2022/reviews/rpp2022-rev-cosmic-rays.pdf}
}

@techreport{HagmannCRY2012,
  author      = {Hagmann, Chris and Lange, David and Verbeke, Jerome and Wright, Doug},
  title       = {{Cosmic-ray Shower Library (CRY)}},
  institution = {Lawrence Livermore National Laboratory},
  year        = {2012},
  month       = mar,
  number      = {UCRL-TM-229453},
  url         = {https://nuclear.llnl.gov/simulation/doc_cry_v1.7/cry.pdf}
}

@manual{radiacode-manual,
  title        = {Radiacode-10X Device Manual},
  organization = {Radiacode},
  year         = {2023},
  url          = {https://downloads.radiacode.com/EN/RC-10x_Device_Manual.pdf},
  note         = {Accessed 2025-08-15}
}

@article{wittmer_properties_1985,
	title = {Properties and microelectronic applications of thin films of refractory metal nitrides},
	volume = {3},
	issn = {0734-2101},
	url = {https://doi.org/10.1116/1.573382},
	doi = {10.1116/1.573382},
	number = {4},
	urldate = {2026-10-13},
	journal = {Journal of Vacuum Science \& Technology A},
	author = {Wittmer, Marc},
	month = jul,
	year = {1985},
	pages = {1797--1803},
}

@article{sherman_growth_1990,
	title = {Growth and {Properties} of {LPCVD} {Titanium} {Nitride} as a {Diffusion} {Barrier} for {Silicon} {Device} {Technology}},
	volume = {137},
	issn = {1945-7111},
	url = {https://iopscience.iop.org/article/10.1149/1.2086826},
	doi = {10.1149/1.2086826},
	language = {en},
	number = {6},
	urldate = {2026-10-13},
	journal = {Journal of The Electrochemical Society},
	publisher = {IOP Publishing},
	author = {Sherman, A.},
	month = jun,
	year = {1990},
	pages = {1892},
}

\FloatBarrier
\appendix

\section{Lower and upper operating voltage from I--V curves \label{app:vbdandul}}
There are several ways to define and measure the breakdown voltage $V_\mathrm{BD}$ of a SiPM. The Onsemi application note~\cite{onsemi_and9770} defines $V_\mathrm{BD}$ as the voltage intercept of a straight-line fit to the linearly increasing region of $\sqrt{I}$ versus $V$. We instead take $V_\mathrm{BD}$ as the inflection point of the $I$--$V$ curve on a logarithmic scale. We fit a Landau--Gaussian convolution function to $\mathrm{d}\ln I/\mathrm{d}V$ and identify the voltage that corresponds to the maximum of this value as $V_{\mathrm{BD}}$. The uncertainly in the maximum of the fit is taken as the uncertainty in $V_{\mathrm{BD}}$. It is necessary to illuminate the SiPM with a constant light source, especially at lower temperatures (below $\sim$ \qty{200}{K}), in order to initiate avalanche above $V_{\mathrm{BD}}$ and measure a rise in current. The upper limit voltage $V_\mathrm{UL}$ is identified from $I$--$V$ curves in which the current saturates, producing a sharp jump at $V_\mathrm{UL}$. Its uncertainty is the voltage step size. $V_{\mathrm{UL}}$ measurements are conducted without illuminating the SiPM.

\section{SiPM characterization from room temperature to \qty{46}{\kelvin}\label{sec:SiPM_characterization}}
This appendix describes the characterization of one MicroFC-60035 SiPM between \qty{293.5}{\kelvin} and \qty{46}{\kelvin}. The SiPM was illuminated with short and weak LED pulses and the amplitude of each output pulse was recorded. A histogram of these amplitudes shows separate peaks for $0, 1, 2, \dots$ detected photons and is referred to as a \emph{finger plot}. From the finger plots we extract the voltage gain, breakdown voltage, optical crosstalk probability, and photon detection efficiency (PDE) relative to its room-temperature value. Since all these quantities come from the same data set, their temperature dependence can be compared directly. The lowest temperature, \qty{46}{\kelvin}, is the coldest the \emph{Attocube} reached with the LED (placed outside the cryostat) shining on the SiPM through the optical window. The breakdown and upper-limit voltages down to millikelvin temperatures were measured separately using $I$--$V$ curves (\autoref{sec:VBD_VUL_vs_T}, method in Appendix \ref{app:vbdandul}).

\subsection{Setup and data set \label{appS:setup}}
The measurements were made in the \emph{Attocube} cryostat (see~\autoref{tab:cryostats}). The sample temperature was read from a Lake Shore DT-670 silicon diode mounted on the sample PCB and driven at \qty{10}{\micro\ampere} by a Keithley~2450 SMU. All temperatures below refer to readings from this diode. The cryostats built-in sensor reads up to \qty{30}{\kelvin} lower than the Lake Shore diode at cooler temperatures, so it was not used. This is believed to be because of insufficient thermalization between the sample and the cold head, worsened by the decrease in cooling power at colder temperatures. The SiPM was biased by a Keysight EDU36311A supply through a single-channel version of the readout box shown in \autoref{fig:readoutboxcircuit}. The output of the readout box was amplified using a low-noise RF amplifier (Mini-Circuits ZX60-P103LN+, \qty{5}{\volt} supply) before being recorded by a Teledyne LeCroy WavePro~254HD oscilloscope with a \qty{20}{\mega\hertz} bandwidth limit.

\begin{figure}[tb]
    \centering
    \includegraphics[width=1\linewidth]{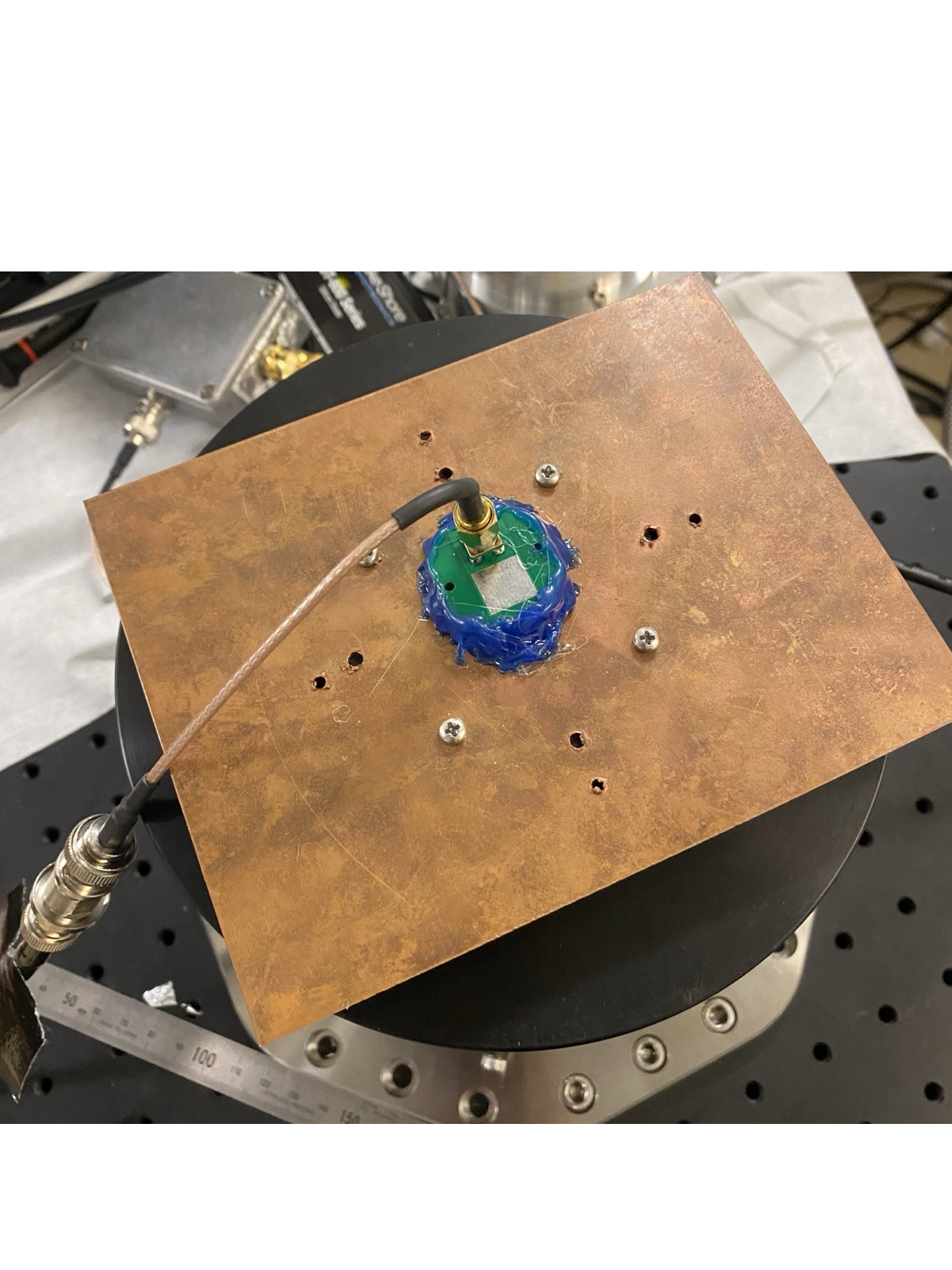}
    \caption{LED assembly for the \emph{Attocube} measurements. The LED is mounted on a small PCB (center), fixed to a copper plate above the optical window of the cryostat. Its light reaches the SiPM through a pinhole.}
    \label{fig:appS_led}
\end{figure}

A blue LED outside the cryostat illuminated the SiPM through a pinhole (see \autoref{fig:appS_led}). It was driven by an Agilent 33220A pulse generator with \qty{40}{\nano\second} pulses at a rate of \qty{1}{\kilo\hertz}. The same generator triggered the oscilloscope, such that every recorded event corresponds to one LED pulse, whether or not a photon was detected. For each event the oscilloscope measured the peak-to-peak amplitude of the SiPM signal in a \qty{46}{\nano\second} gate containing the pulse maximum. Measuring in a narrow window and relative to the local baseline suppresses pile-up and slow baseline drifts. The idea follows ~\citet{Rehbein2023}, where SiPM pulses were integrated over only \qty{15}{\nano\second} after subtracting a local baseline taken just before the leading edge. These amplitudes were histogrammed on the oscilloscope (200 bins). Each histogram contains about \num{20000} LED pulses.

\begin{figure*}[tb]
    \centering
    \includegraphics[width=\textwidth]{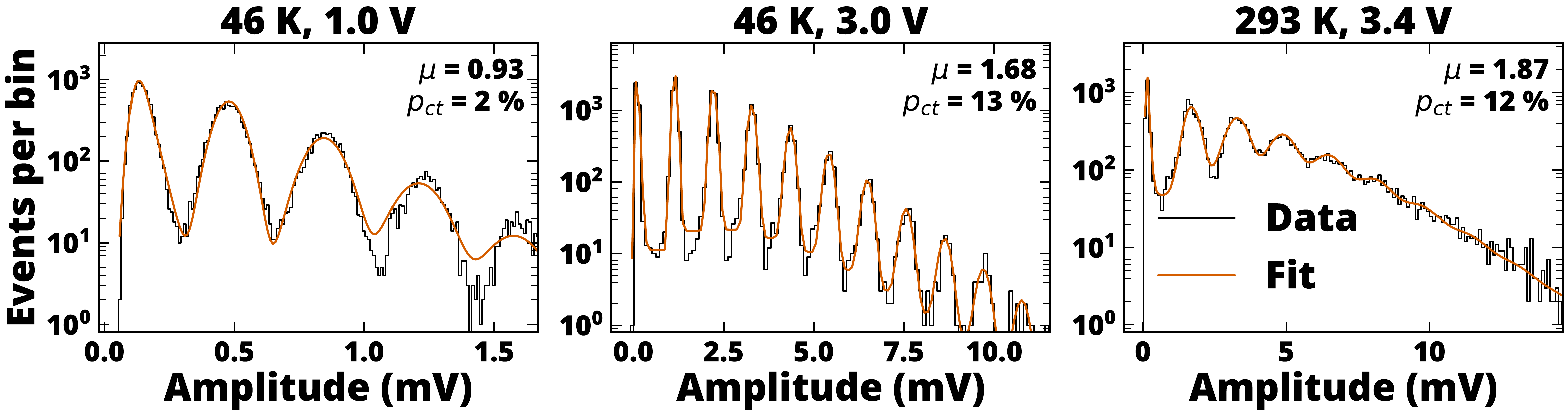}
    \caption{Example finger plots (peak-to-peak amplitude per LED pulse) with the full-histogram fit of Appendix \ref{appS:model}.
    Left: \qty{46}{\kelvin}, \qty{1.0}{\volt} overvoltage: well-resolved peaks.
    Middle: \qty{46}{\kelvin}, \qty{3.0}{\volt}: the peaks are narrower.
    Right: \qty{293}{\kelvin}, \qty{3.4}{\volt}: wider, asymmetric peaks and a continuum between them.
    $\mu$ is the mean number of detected photons per pulse and $p_\mathrm{ct}$ the crosstalk probability.}
    \label{fig:appS_fingers}
\end{figure*}

The temperature campaign covered 16 temperatures from \qty{46.3}{\kelvin} to \qty{293.5}{\kelvin}, with an automated scan at each. The scans applied five bias voltage values between about \qty{1}{\volt} and \qty{5}{\volt} overvoltage (up to \qty{4}{\volt} at \qty{270.8}{\kelvin} and \qty{293.5}{\kelvin}). In total 80 finger plots were acquired. During each scan the temperature changed by less than \qty{0.3}{\kelvin}. The LED pulse settings were identical at all 80 points. The LED stayed at room temperature, so the light reaching the SiPM is assumed to be the same throughout. Representative finger plots are shown in \autoref{fig:appS_fingers}.

\subsection{Gain and breakdown voltage \label{appS:gain}}
When a SiPM is operated above the breakdown voltage, a single detected photon normally triggers an avalanche in one microcell, which is stopped by the quenching resistor of that cell. The number of charge carriers released per avalanche is the gain~\cite{onsemi_and9770},
\begin{equation}
    \label{eq:appS_gain_cj}
    G=\frac{C_{\mathrm{J}}\,V_\mathrm{OV}}{q_{\mathrm{e}}},
\end{equation}
where $C_{\mathrm{J}}$ is the junction capacitance of a microcell, $V_\mathrm{OV}$ is the overvoltage and $q_{\mathrm{e}}$ is the electron charge. At a given temperature the gain is therefore linear in overvoltage; at fixed overvoltage it can change with temperature through the temperature dependence of $C_{\mathrm{J}}$. For the MicroFC-60035 the datasheet gives $G \approx 3\times10^{6}$ at \qty{2.5}{\volt} overvoltage at room temperature~\cite{OnsemiMicroC2022}. Each avalanche gives nearly the same signal, so the pulse amplitude is proportional to the number of detected photons, which produces the separate peaks of the finger plot (labeled $0, 1, 2,\dots$)  photoelectrons (p.e.) in \autoref{fig:appS_fingers}.

We measure the gain as the spacing $\Delta V$ between neighboring peaks. We record a peak-to-peak voltage behind an amplifier, not a charge. The amplifier is only rated for 50 and \qty{3000}{\mega\hertz}, whereas our signals are limited to below \qty{20}{\mega\hertz}. We therefore cannot convert the amplitude into a charge. The voltage gain we quote describes the whole readout chain rather than the SiPM alone, but it tracks relative changes with temperature and overvoltage. At lower temperatures the dark count rate is negligible. At room temperature, dark pulses fall in 5--\qty{21}{\percent} of the gates (Appendix \ref{appS:dark}), but with the short LED-triggered gate the peaks are still resolved (see \autoref{fig:appS_fingers}, right).
The spacing is extracted with an FFT of the histogram, which gives a starting value for the Gaussian analysis. After subtracting a smooth background, Gaussians are fit to the individual peaks to give the final spacing. The cross-check in Appendix \ref{appS:model} uses an independent fit of the whole histogram. The two methods agree to about \qty{1}{\percent} above \qty{2}{\volt} overvoltage.

\begin{figure}[tb]
    \centering
    \includegraphics[width=\linewidth]{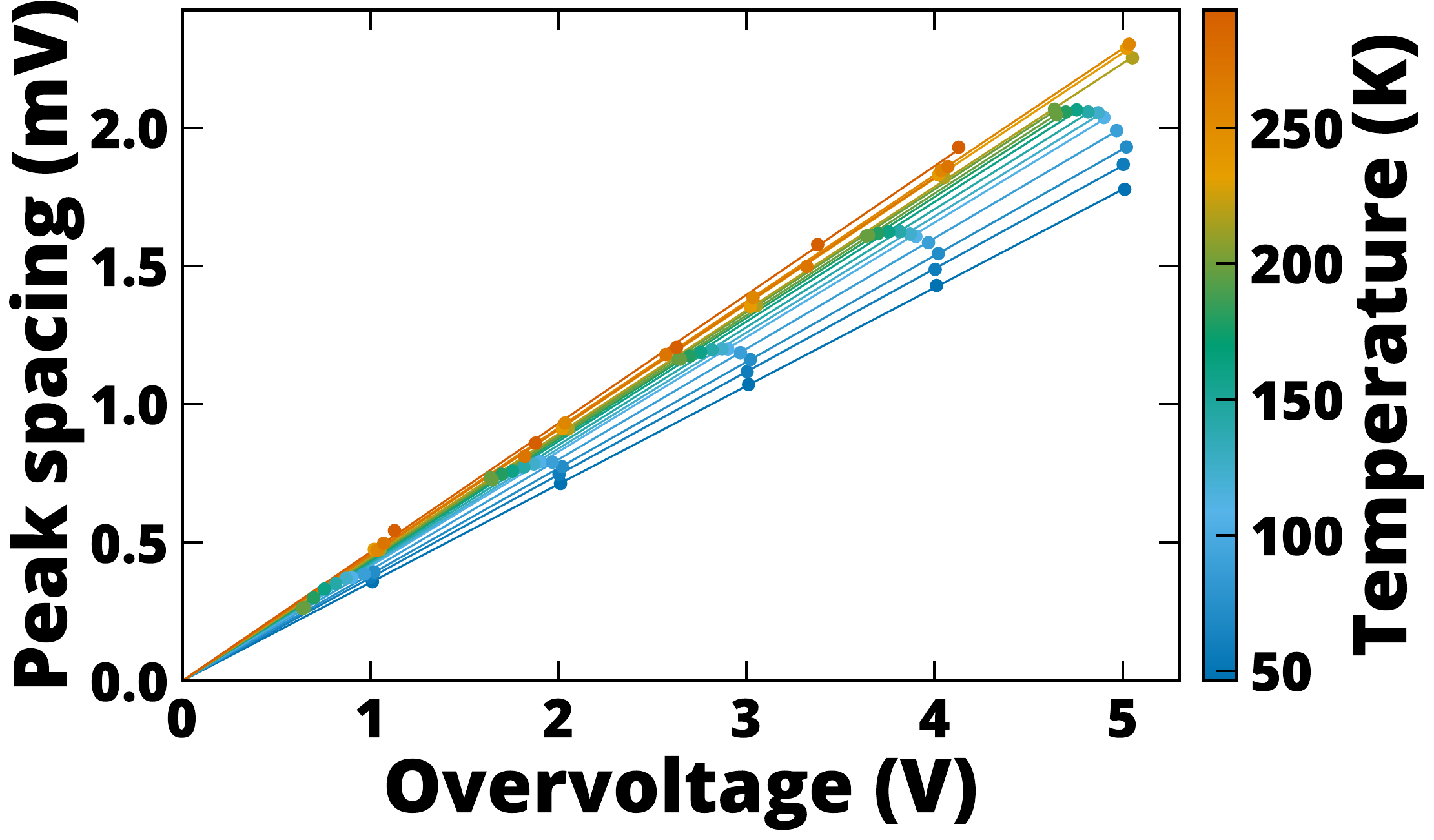}\\
    \includegraphics[width=\linewidth]{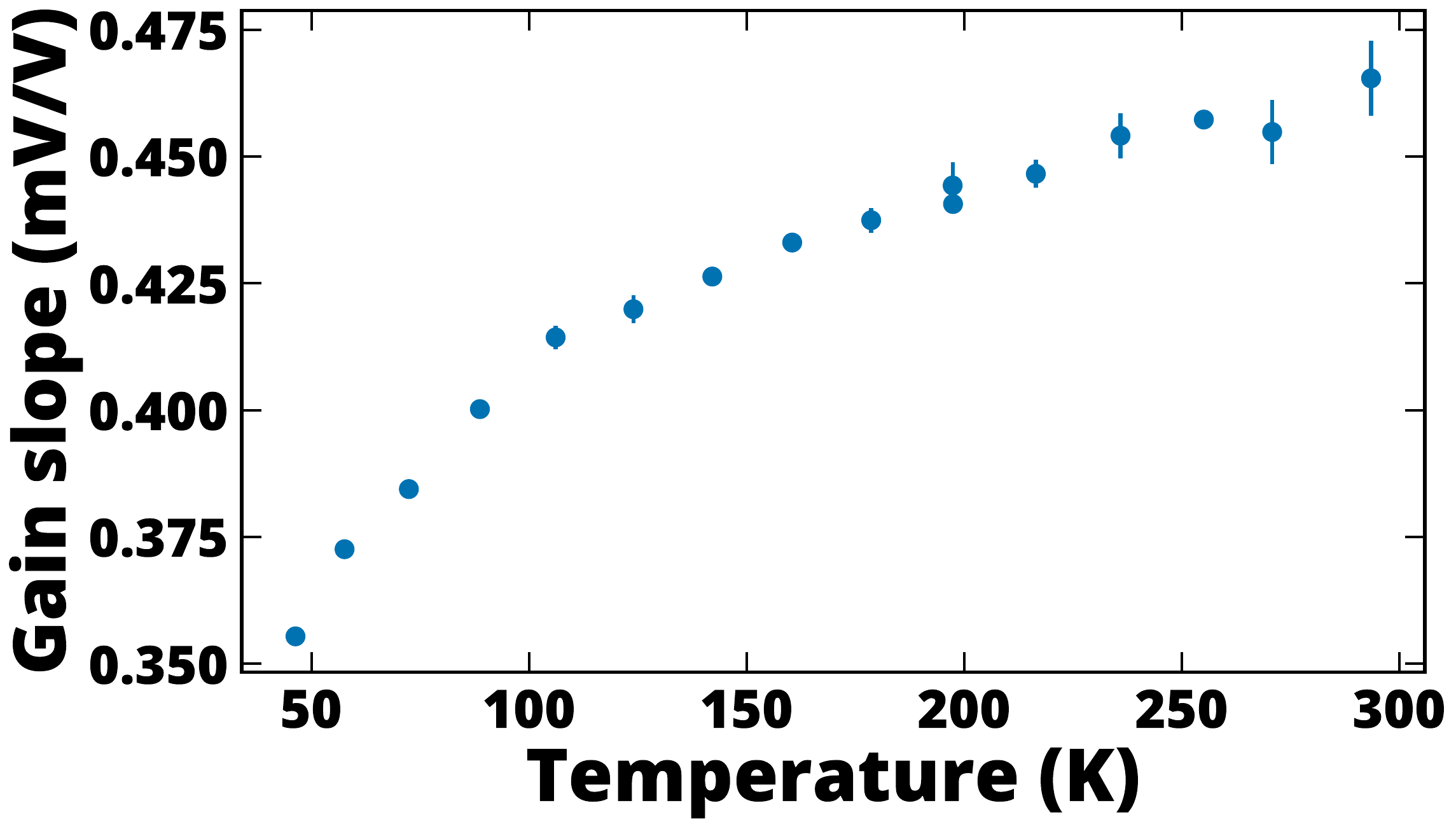}
    \caption{Top: photoelectron peak spacing against overvoltage at 16 temperatures. Lines are linear fits; their zero crossing defines $V_\mathrm{BD}$.
    Bottom: slope of the fits against temperature.}
    \label{fig:appS_gain}
\end{figure}

As expected from Eq.~\eqref{eq:appS_gain_cj}, at each temperature $\Delta V$ is linear in the bias voltage (see \autoref{fig:appS_gain}, top; $R^2 > 0.999$ for every scan). The slope $\mathrm{d}\Delta V/\mathrm{d}V$ is the gain per volt, and the bias at which the line extrapolates to zero spacing is the breakdown voltage $V_\mathrm{BD}$. This value is used to compute the overvoltage throughout this appendix. The fits are robust: removing any single bias point changes $V_\mathrm{BD}$ by less than \qty{0.1}{\volt}. $V_\mathrm{BD}$ rises by \qty{3.95}{\volt} from \qty{46}{\kelvin} to \qty{293.5}{\kelvin}, with $\mathrm{d}V_\mathrm{BD}/\mathrm{d}T \approx \qty{20}{\milli\volt\per\kelvin}$ near room temperature. The room-temperature value, $V_\mathrm{BD}=\qty{24.67(5)}{\volt}$ at \qty{20.3}{\celsius}, lies inside the manufacturer's range of 24.2--\qty{24.7}{\volt} at \qty{21}{\celsius}~\cite{OnsemiMicroC2022}.

The gain-intercept $V_\mathrm{BD}$ need not equal the $I$--$V$ value of \autoref{sec:VBD_VUL_vs_T}. At \qty{44.85}{\kelvin}, for example, the gain intercept is \qty{0.16}{\volt} above the $I$--$V$ inflection point. Different $I$--$V$ definitions of $V_\mathrm{BD}$ alone differ by up to \qty{0.3}{\volt}.

The slope $\mathrm{d}\Delta V/\mathrm{d}V$ grows by \qty{31}{\percent} between \qty{46}{\kelvin} and \qty{293.5}{\kelvin} (see \autoref{fig:appS_gain}, bottom). Because we record a band-limited amplitude rather than a charge, this change cannot be attributed to $C_J$ alone, and may come from the pulse shape rather than the capacitance.

\subsection{Photon statistics and crosstalk model \label{appS:model}}

The number of photons detected per LED pulse (``primary'' avalanches) follows a Poisson distribution with mean $\mu$. Optical crosstalk results in additional avalanches. A fraction of the carriers in an avalanche emits photons, which can fire neighboring microcells at the same time. In the finger plot, each LED pulse is counted in the peak matching its number of avalanches. Crosstalk adds avalanches to a pulse that already has at least one, so it moves that pulse from the 1~p.e.\ peak to the 2~p.e.\ peak or higher. A pulse with no detected photon has no avalanche to start crosstalk, so the 0~p.e.\ peak is unaffected. Hence, for any crosstalk model,
\begin{align}
P(0) &= e^{-\mu}, & P(1) &= \mu\,e^{-\mu}\,(1-p_\mathrm{ct}),
\label{eq:appS_p01}
\end{align}
where $p_\mathrm{ct}$ is the probability that an avalanche fires at least one neighbor~\cite{Gallego2013,eckertCharacterisationStudiesSilicon2010}. Counting the events $N_0$ and $N_1$ in the 0 and 1~p.e.\ peaks out of $N$ LED pulses gives model-independent estimators,
\begin{align}
\mu_0 &= -\ln\frac{N_0}{N}, & p_\mathrm{ct} &= 1-\frac{N_1}{\mu_0 N_0}.
\label{eq:appS_estimators}
\end{align}
A plain Poisson fit (ignoring crosstalk) to all peaks overestimates $\mu$ by up to \qty{35}{\percent} at \qty{5}{\volt}, because crosstalk shifts events to higher peaks.

The higher peaks depend on how many neighbors one avalanche can fire. We take this number to be Poisson-distributed with mean $\lambda$, which gives the generalized Poisson distribution~\cite{Vinogradov2012,Chmill2017},
\begin{equation}
P(k)=\frac{\mu\,(\mu+k\lambda)^{k-1}e^{-(\mu+k\lambda)}}{k!}, \qquad p_\mathrm{ct}=1-e^{-\lambda}.
\label{eq:appS_gp}
\end{equation}
We fixed $\mu$ and $p_\mathrm{ct}$ from Eq.~\eqref{eq:appS_estimators} and predicted the 2--4~p.e.\ peaks. The generalized Poisson model predicts them to about \qty{1}{\percent}, while plain Poisson is off by up to \qty{23}{\percent}.

We then fitted each histogram as a whole with Eq.~\eqref{eq:appS_gp}, using a binned Poisson likelihood. Besides $\mu$ and $p_\mathrm{ct}$, the fit has free parameters for the pedestal position, the gain and the peak widths, and a continuum for pulses that only partly overlap with the gate (dark pulses, delayed crosstalk and afterpulses), which add between 0 and 1~p.e. The model reproduces all 80 histograms (\autoref{fig:appS_fingers}). Unless noted otherwise, all values of $\mu$ and $p_\mathrm{ct}$ below come from this full fit.

Dark pulses complicate the simple estimators at room temperature. A dark pulse that only partly overlaps with the gate adds less than 1~p.e., so it moves an event with no detected photon out of the zero peak and into the 1~p.e.\ window. This makes $\mu_0$ too large and pushes $p_\mathrm{ct}$ from Eq.~\eqref{eq:appS_estimators} to almost zero at \qty{293.5}{\kelvin}. The full fit assigns these events to the continuum and keeps them in the correct class.

We checked this with a room-temperature run taken with the LED disconnected (``LED off''), which records only dark pulses. Dark pulses are independent of the LED light, so the probability of an empty gate with the LED on is the product of the two, $N_0^\mathrm{on}/N^\mathrm{on} = e^{-\mu}\,N_0^\mathrm{off}/N^\mathrm{off}$. This gives a dark-corrected estimate,
\begin{equation}
\mu_\mathrm{dark} = -\ln\frac{N_0^\mathrm{on}}{N^\mathrm{on}} + \ln\frac{N_0^\mathrm{off}}{N^\mathrm{off}},
\label{eq:appS_mu_dark}
\end{equation}
which needs no crosstalk or continuum model~\cite{eckertCharacterisationStudiesSilicon2010,Klanner2019}. The full-fit $\mu$ agrees with it to 1--\qty{2}{\percent}.

\subsection{Optical crosstalk \label{appS:crosstalk}}
At each temperature the crosstalk probability rises with overvoltage, from about \qty{2}{\percent} at \qty{1}{\volt} to 15--\qty{25}{\percent} at \qty{5}{\volt}. The scans at different temperatures were taken at slightly different overvoltages. To compare them at fixed overvoltage, we interpolate the crosstalk of each scan linearly between the two neighboring overvoltages.

\begin{figure}[tb]
    \centering
    \includegraphics[width=\linewidth]{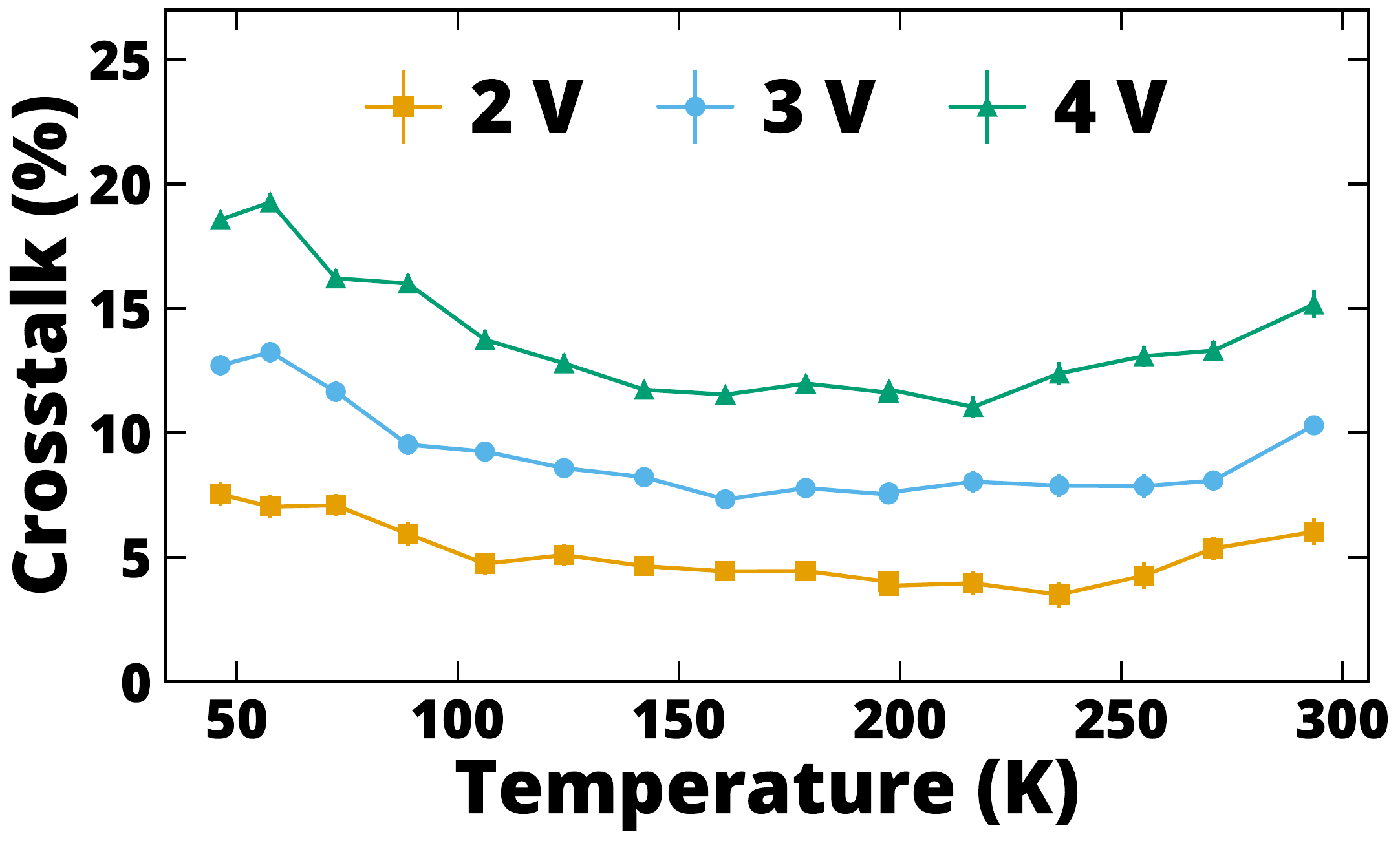}
    \caption{Optical crosstalk probability at fixed overvoltage against temperature, from the full-histogram fit, linearly interpolated between the measured overvoltages of each scan. Error bars are statistical.}
    \label{fig:appS_crosstalk}
\end{figure}

The crosstalk is highest at the coldest temperatures (see \autoref{fig:appS_crosstalk}). At \qty{3}{\volt} overvoltage it is about \qty{13}{\percent} at 46--\qty{58}{\kelvin}, falls to a nearly constant \qty{8}{\percent} between about \qty{125}{\kelvin} and \qty{271}{\kelvin}, and is \qty{10.3}{\percent} at \qty{293.5}{\kelvin}. The gain cannot explain the rise at low temperatures: at fixed overvoltage the gain per volt is about \qty{24}{\percent} lower at \qty{46}{\kelvin} than at room temperature (see \autoref{fig:appS_gain}, bottom), which would lower the crosstalk. The crosstalk per avalanche carrier is therefore larger at lower temperatures. At room temperature and \qty{2.5}{\volt} we find $p_\mathrm{ct}=\qty{8.3}{\percent}$, compared with the typical datasheet value of \qty{7}{\percent}~\cite{OnsemiMicroC2022}. The two scans at \qty{197}{\kelvin}, taken an hour apart, agree within the statistical errors shown.

\subsection{Photon detection efficiency relative to room temperature \label{appS:pde}}
\begin{figure}
    \centering
    \includegraphics[width=\linewidth]{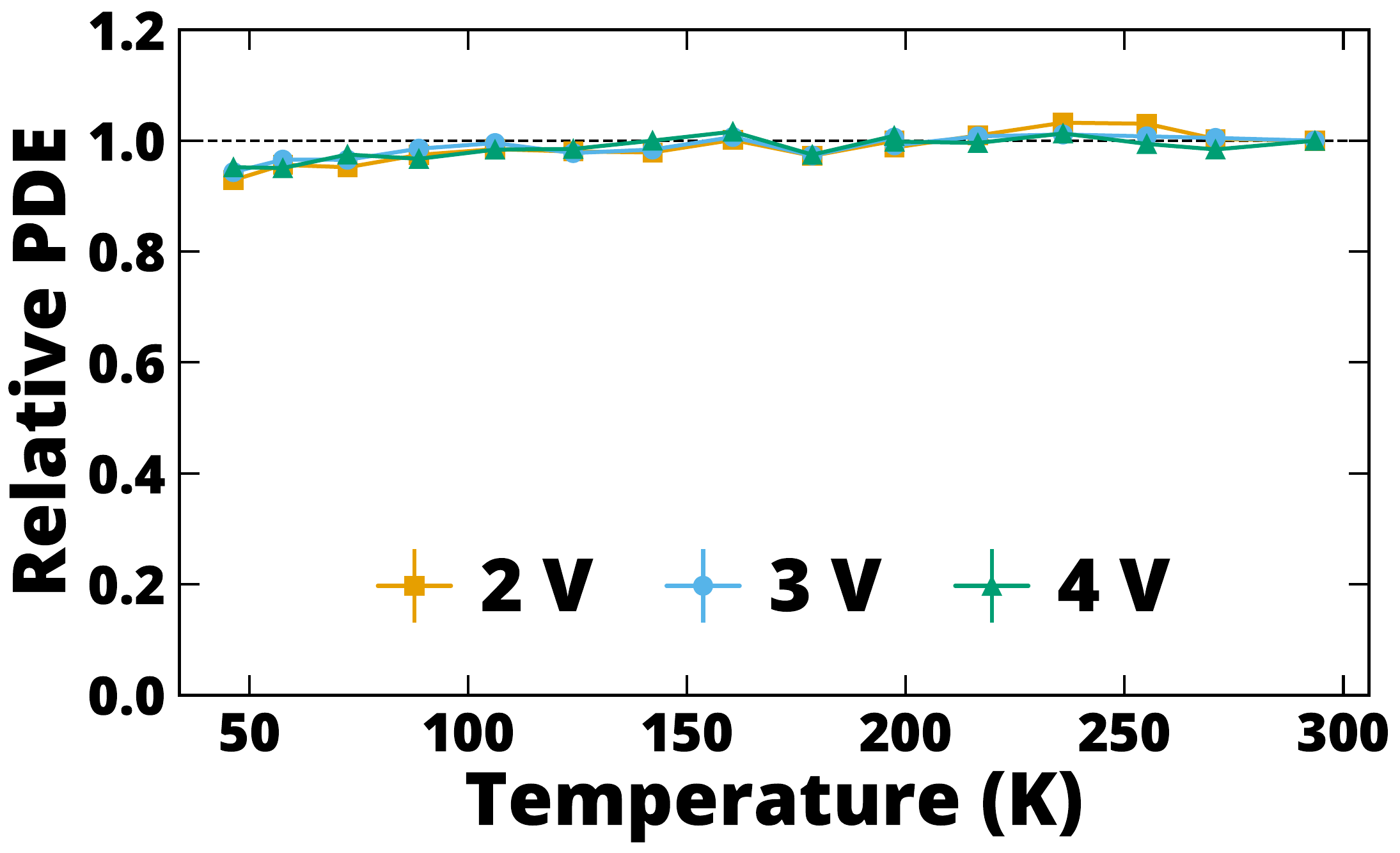}
    \caption{PDE relative to \qty{293.5}{\kelvin} at fixed overvoltage, from the mean number of detected photons per LED pulse. The error bars are statistical only; the scatter between repeated scans is 2--\qty{3}{\percent}.}
    \label{fig:appS_pde}
\end{figure}
The PDE is the fraction of incident photons that are detected. It depends on the quantum efficiency of the microcells, the avalanche triggering probability, and the geometric fill factor of the microcell array. An absolute PDE measurement requires a stable light source of known power. Lacking one, we measure the PDE relative to room temperature. The LED illumination is the same at every point, so $\mu$ is proportional to the PDE. Crosstalk does not change $\mu$, because $\mu$ counts primary avalanches only. This matches the datasheet definition of the PDE, which excludes crosstalk and afterpulsing~\cite{OnsemiMicroC2022}. Unlike a PDE estimate from the integrated photocurrent, $\mu$ does not depend on the gain. To compare temperatures, the values of $\mu$ from each scan were interpolated to fixed overvoltages of 2, 3 and \qty{4}{\volt}. They were then normalized to the room-temperature scan, where the PDE is known from the datasheet (see \autoref{fig:appS_pde}). We leave out \qty{1}{\volt} because the room-temperature reference there is still affected by dark pulses.

Between about \qty{150}{\kelvin} and room temperature the PDE is constant within about \qty{2}{\percent}. It drops to 0.93--0.95 of the room-temperature value at \qty{46}{\kelvin}. 

We can calculate a proxy of absolute PDE by referencing our relative values to the room temperature value given on the data sheet for PDE at \qty{420}{\nano\meter}.

 In terms of absolute PDE, this then corresponds to a fall from \qty{34.6}{\percent} to \qty{32.6}{\percent} at \qty{3}{\volt}. Repeated scans scatter by 2--\qty{3}{\percent}, more than their statistical errors. The scans were taken in order of increasing temperature over 2.5 days, so a slow drift of the LED or of the light path would look like a temperature dependence; we have not yet measured such a drift independently.

\begin{figure*}
    \centering
    \includegraphics[width=0.8\linewidth]{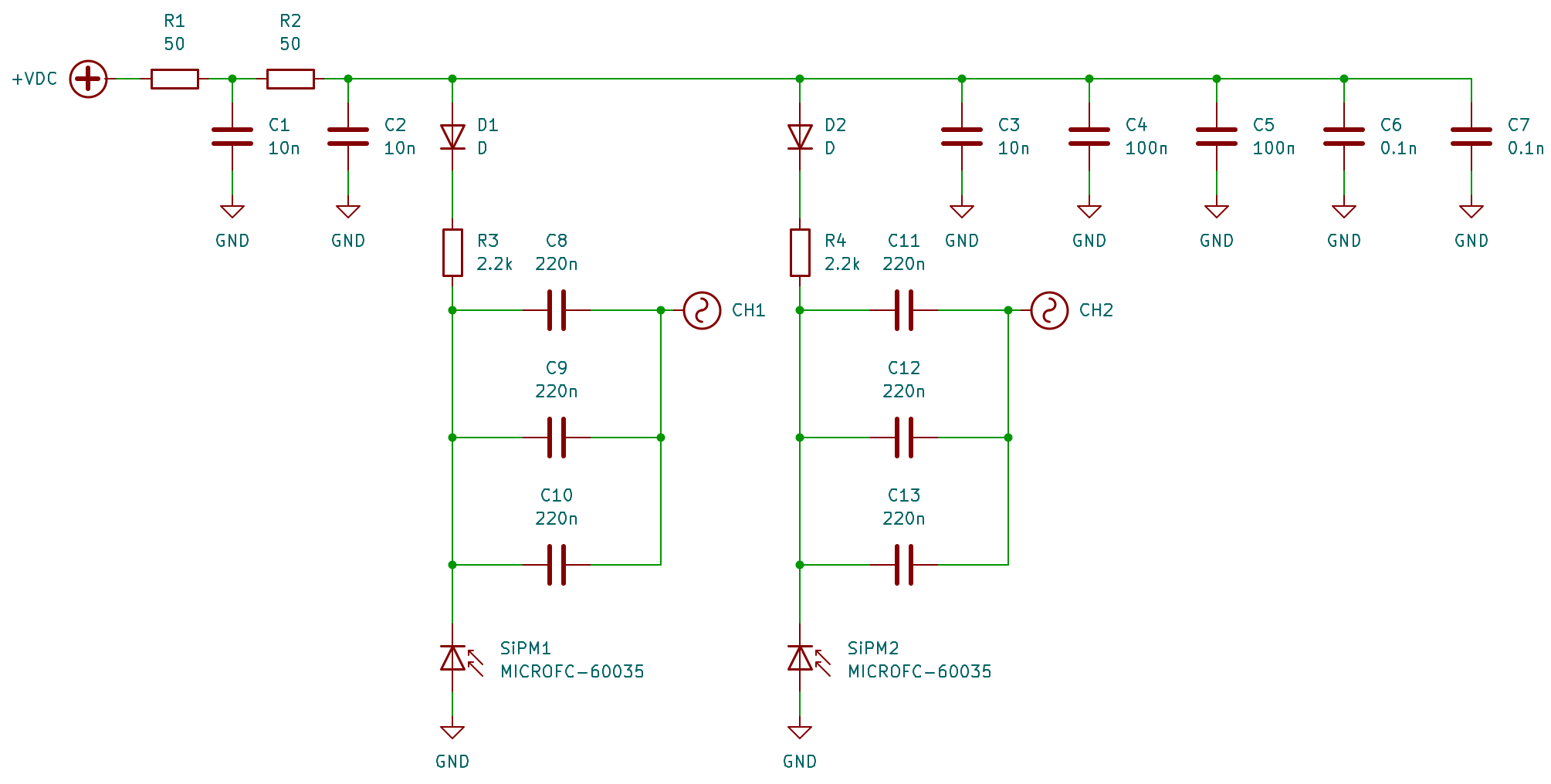}
    \caption{Circuit diagram of readout box. }
    \label{fig:readoutboxcircuit}
\end{figure*}
At room temperature, the overvoltage dependence of $\mu_\mathrm{dark}$ from Eq.~\eqref{eq:appS_mu_dark} follows the datasheet PDE curve to about \qty{1}{\percent} between \qty{1.1}{V} and \qty{4.1}{\volt}. If the LED emitted at \qty{420}{\nano\meter}, this would correspond to $5.2$ photons per pulse on the SiPM. Our LED emits at about \qty{465}{\nano\meter}, where the datasheet PDE is lower, and its spectrum was not measured, so we do not quote an absolute PDE.

\subsection{Finger-plot resolution \label{appS:resolution}}
\begin{figure}[tb]
    \centering
    \includegraphics[width=\linewidth]{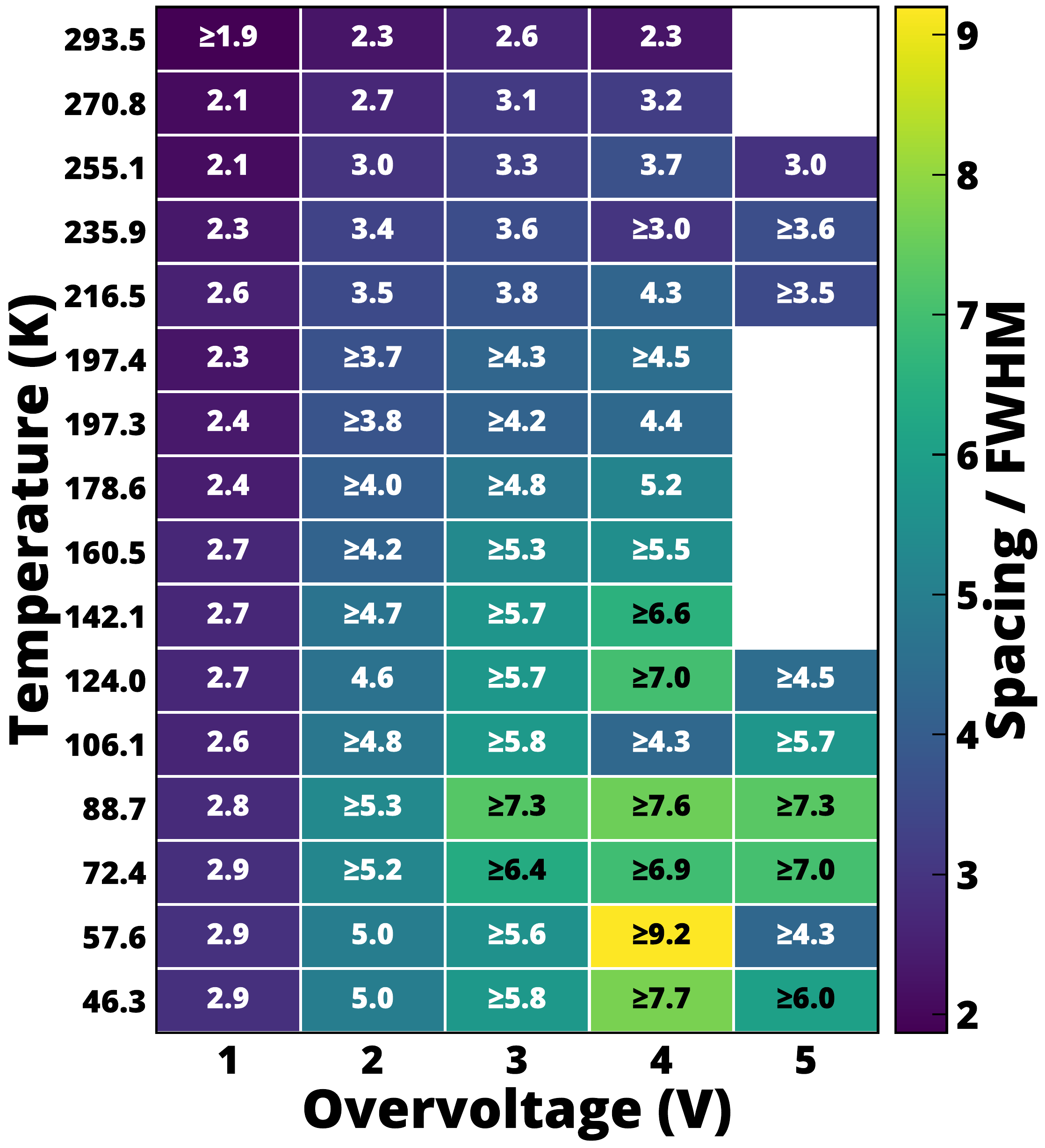}
    \caption{Finger-plot resolution, the peak spacing divided by the FWHM of the 1~p.e.\ peak, at fixed overvoltage for each scan. ``$\geq$'' marks lower limits, where the 1~p.e.\ peak is narrower than three histogram bins. Blank cells are overvoltages a scan did not reach. Two scans were taken at \qty{197}{\kelvin}.}
    \label{fig:appS_resolution}
\end{figure}
We define the resolution as the spacing between neighboring peaks divided by the full width at half maximum (FWHM) of the 1~p.e.\ peak. The FWHM is read directly from the histogram, without fitting a peak shape: starting from the highest bin near the 1~p.e.\ position, we find where the counts fall below half of that height on each side, interpolating linearly between bins. Values at fixed overvoltage are taken from the bias point ($V\mathrm{OV} = 1, 2, \dots)$ within \qty{0.15}{\volt}, or else interpolated between the two neighboring points. The resolution improves with overvoltage and at low temperature (\autoref{fig:appS_resolution}). At \qty{2}{\volt} it rises from 2.3 at \qty{293.5}{\kelvin} to 5.0 at \qty{46}{\kelvin}. At $\ge$ \qty{2}{\volt}, and below about \qty{200}{\kelvin}, most 1~p.e.\ peaks are narrower than three of the 200 histogram bins, so only lower limits can be given.

\subsection{Dark counts \label{appS:dark}}
At room temperature, the LED-off run gives the fraction of gated events containing a dark pulse. This fraction rises from \qty{5}{\percent} at \qty{1.1}{\volt} to \qty{21}{\percent} at \qty{4.1}{\volt} overvoltage (\qty{294.5}{\kelvin}). At the cold end of the campaign, dark pulses are negligible for the measurements in this appendix. At \qty{8.5}{\milli\kelvin} in \emph{Bluefors} we counted 17 dark pulses above a \qty{0.94}{\milli\volt} threshold in \qty{422}{\second} at \qty{24.5}{\volt} bias. This corresponds to a dark count rate of \qty{0.04}{\hertz}, compared with the typical \qty{1.2}{\mega\hertz} at room temperature for this device~\cite{OnsemiMicroC2022}.
\section{Readout Box/Electronics}

A schematic of the electronics inside the readout box is given in \autoref{fig:readoutboxcircuit}. R1, R2, C1, C2 constitute a low-pass RC filter for the bias voltage. C3-C7 are shunt capacitors to remove high-frequency noise. D1 and D2 prevent one channel from drawing current from the other, which would cause undershoot and ringing. R3, R4 ensure the path from the SiPM to the voltage supply has a higher impedance than the path from the SiPM to the oscilloscope. C8-C10 and C11-C13 are for decoupling the output signals from the cathode lines. The oscilloscope inputs were terminated with external \qty{50}{\ohm} terminations. 

\section{Simulation \label{sec:Simulation}}
We measured the muon count rate at temperatures from room temperature down to millikelvin and compared it with the rate predicted by a Monte Carlo simulation in Geant4 (see \autoref{sec:muon_rate}).

The simulated detector consists of two plastic scintillator blocks wrapped in ESR film, each with a SiPM coupled to one end, positioned inside a simplified dilution refrigerator geometry consisting of three concentric cans and the base plate of the relevant stage. We expect the omitted material, such as intermediate plates and wiring, to have a negligible effect on the simulated muon rate.

We simulated atmospheric muons using the CRY package~\cite{HagmannCRY2012} and scaled the flux by an attenuation factor of \qty{26}{\percent}, measured by comparing the muon flux outside of the building with that in the room of the cryostat.

We also measured the gamma spectrum inside the laboratory right next to the dilution fridge with a Radiacode-103 detector~\cite{radiacode-manual}, and encoded it as a gamma source in a general particle source enclosing the fridge geometry. We recorded the energy of photon hits on the SiPMs and scaled the number of photons using the wavelength dependence of the PDE reported by Onsemi for the MicroFC-30035 (same \qty{35}{\micro\metre} microcell) at room temperature~\cite{OnsemiMicroC2022}. Photon counts were converted to voltage signals using a gain factor of \qty{1.5}{\milli\volt} per photon. Muons were identified as coincident events in both SiPMs, where scintillation in both blocks was caused by the same muon, and a threshold of \qty{23}{\milli\volt} was applied to count muon hits. This threshold is consistent with the threshold applied in \autoref{sec:analytical_rate}. The coincidence rate was obtained by converting the number of simulated muons into an exposure time, using the sea-level rate of \qty{1}{\per\square\centi\metre\per\minute} through a \qtyproduct{20 x 20}{\centi\metre} slab above the refrigerator. All scintillation photons in the simulation were produced by muons; no gamma interactions with the detector were recorded. Radiation sources inside the dilution refrigerator were not included in the simulation.

\subsection{Analytical coincidence rate \label{sec:analytical_rate}}
The analytical coincidence rate is
\begin{equation}
R = \qty{8.1 \pm 1.4}{\per\hour}.
\end{equation}
We take $I(\theta)=I_0\cos^2\theta$ with $I_0=2\Phi/\pi$, so that the flux through a horizontal surface is $\Phi=\qty{1}{\per\square\centi\metre\per\minute}$~\cite{PDGCosmicRays2022}. The scintillator blocks are stacked, with horizontal $a\times b=\qtyproduct{8 x 14}{\milli\metre}$ faces, height $h=\qty{8}{\milli\metre}$ and gap $g=\qty{17}{\milli\metre}$. A track with a path of at least $L_\mathrm{min}$ in each block crosses the two planes a depth $L_\mathrm{min}\cos\theta$ inside the blocks, which are $s=g+2L_\mathrm{min}\cos\theta$ apart. Seen along the track, these planes are offset by $s\tan\theta$, so the area accepting tracks at $(\theta,\phi)$ is their overlap,
\begin{equation}
A=\big[a-s\tan\theta\,|\cos\phi|\big]_+\,\big[b-s\tan\theta\,|\sin\phi|\big]_+ ,
\end{equation}
where $[x]_+=\max(x,0)$ sets $A=0$ when they do not overlap ($A=0$ also if $L_\mathrm{min}\cos\theta>h$). The rate is
\begin{equation}
R=\frac{2\Phi T}{\pi}\int_0^{2\pi}\!\!\int_0^{\pi/2}\!A\,\cos^3\theta\,\sin\theta\,\mathrm{d}\theta\,\mathrm{d}\phi ,
\end{equation}
with the building transmission $T=0.74$ and $L_\mathrm{min}\approx\qty{0.4}{\milli\metre}$ for the threshold. The error is $\sigma_R=|\partial R/\partial g|\,\sigma_g$ with $\sigma_g=\qty{2}{\milli\metre}$.

\section{Double Pulses \label{sec:doublepulseappendix}}
\subsection{Effects of magnetic field on double pulses \label{subsec:doublepulsesappendixa}}

\begin{figure}
    \centering
    \includegraphics[width=0.8\linewidth]{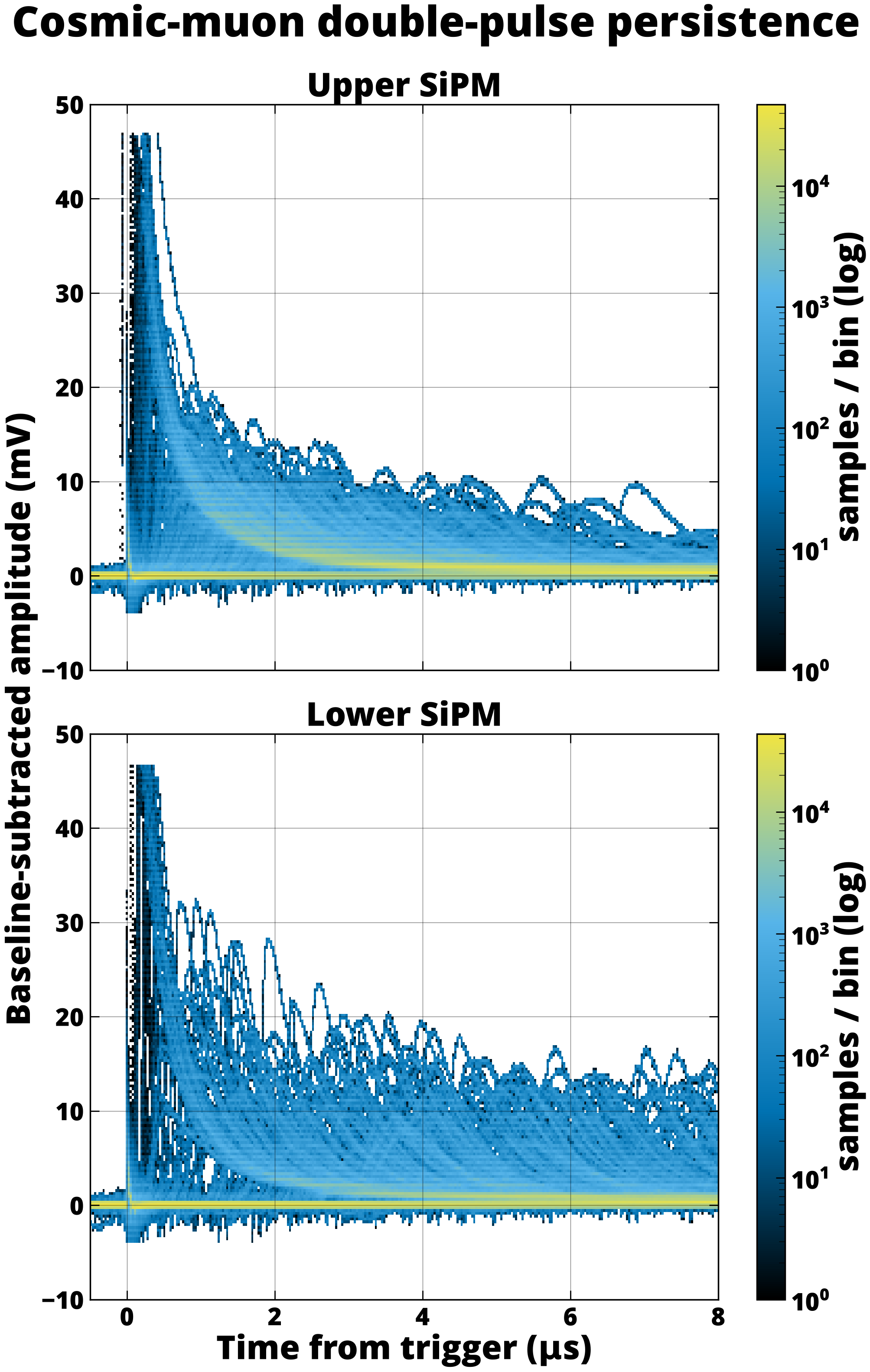}
    \caption{Overlaid double-pulse waveforms from cosmic muons at millikelvin temperatures, for the upper (top) and lower (bottom) SiPMs. The narrow pulses all overlap at the trigger time. The broad pulses vary in amplitude, and their amplitude decreases the further they are in time from the narrow pulse (i.e., the delay and the amplitude of the broad pulse are anticorrelated).}
    \label{fig:persistence}
\end{figure}
Two SiPMs were operated at temperatures between \qty{6}{K} and \qty{8}{K} to check if double pulsing below \qty{8}{K} was caused by some component in the SiPM becoming superconducting. This was achieved by mounting the SiPMs in \emph{Bluefors} as described in \autoref{sec:Double_pulses}. We varied the magnetic field (between 0 and \qty{4.5}{\tesla}) at different temperatures while observing the pulse shape of the SiPM output when triggered by an LED (also mounted on the millikelvin stage). Turning up the magnetic field to a sufficiently high field strength causes the double pulsing behavior to disappear. The field was ramped at \qty{6.02}{K}, \qty{7.00}{K} (field ramped down), \qty{7.04}{K} (field ramped up), \qty{7.53}{K} and \qty{7.75}{K}; half of the pulses were merged at \qty{3.64}{T}, \qty{1.69}{T}, \qty{1.90}{T}, \qty{0.90}{T} and \qty{0.81}{T}, respectively. At \qty{8}{K} the pulses are already merged without an applied field. 
A waveform is called `merged' when the broad component is no longer separated from the narrow one, i.e.\ it starts within the duration of the narrow pulse (about \qty{25}{ns}), giving a single pulse. In practice, merged waveforms are identified by a mean signal above \qty{1}{\milli\volt} between 150 and \qty{700}{ns} after the trigger; this quantity is bimodal, with only about \qty{1}{\percent} of waveforms between the two populations. The proportion of merged pulses to double pulses was plotted as a function of the magnetic field for different temperatures to create a transition graph as shown in the inset of \autoref{fig:pulse_shape_transition}. The field strength at which the transition occurs is defined as the point at which 50\% of the waveforms have merged. The relationship between SiPM temperature and transition field strength is shown in \autoref{fig:pulse_shape_transition}. The fields at which \qty{10}{\percent}, \qty{50}{\percent} and \qty{90}{\percent} of the pulses are merged were each fitted with the parabolic temperature dependence of the critical field,
\begin{equation}
    B(T) = B_0\left[1-\left(\frac{T}{T_0}\right)^{2}\right],
    \label{eq:bc_of_t}
\end{equation}
giving $T_0 = \qty{7.99 \pm 0.08}{K}$, \qty{8.05 \pm 0.08}{K} and \qty{8.10 \pm 0.09}{K}, and $B_0 = \qty{7.3 \pm 0.4}{T}$, \qty{8.1 \pm 0.5}{T} and \qty{8.9 \pm 0.5}{T}, respectively. The \qty{10}{\percent} and \qty{90}{\percent} fits bound the mixed region in \autoref{fig:pulse_shape_transition}; we quote $T_c$ and $B_{c2}(0)$ from the \qty{50}{\percent} fit.

\begin{figure}
    \centering
    \includegraphics[width=\linewidth]{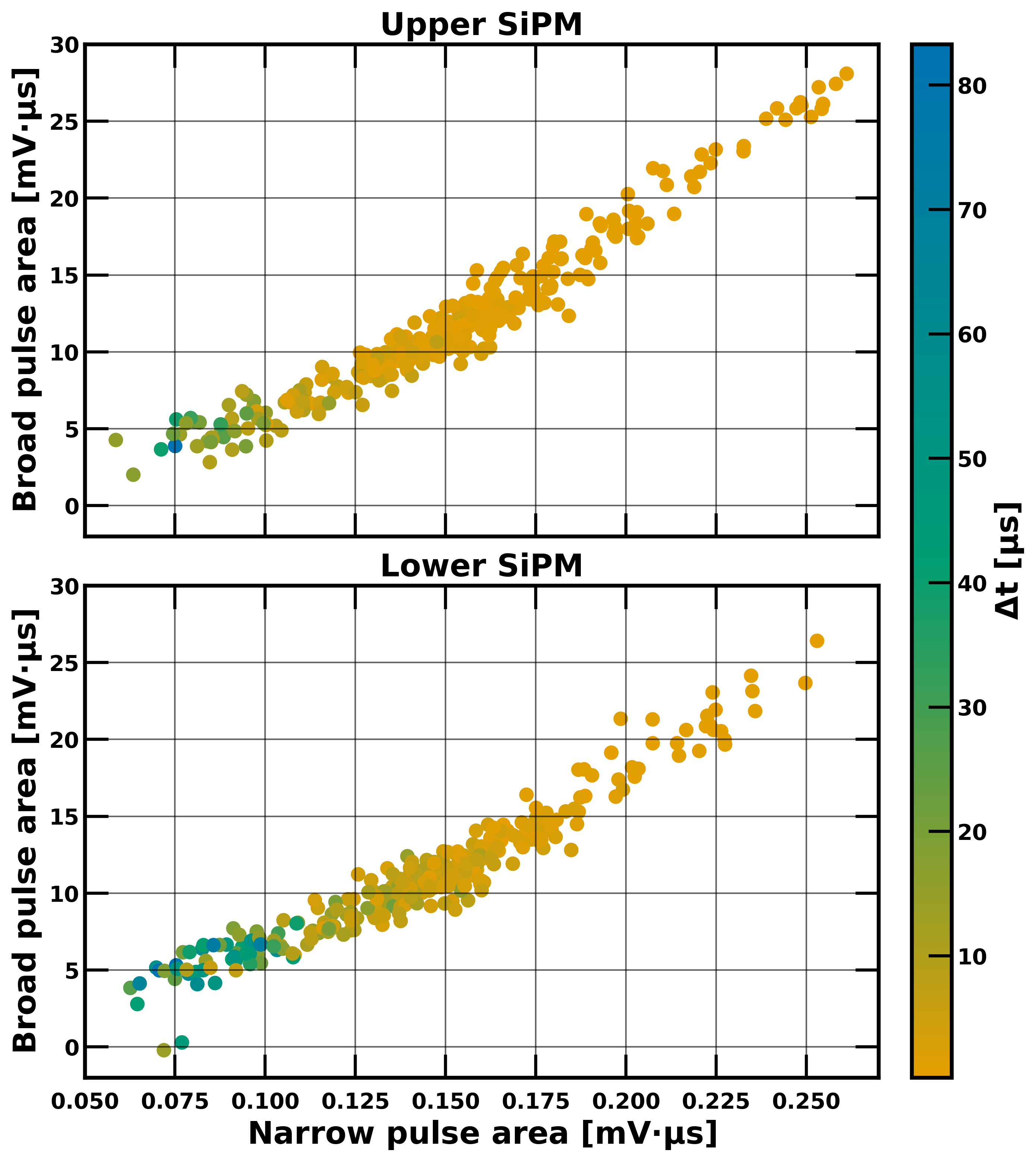}
    \caption{Area of the broad pulse versus area of the narrow pulse for muon events with double pulses, for the upper SiPM (top) and the lower SiPM (bottom). The color scale shows the time $\Delta t$ between the narrow and broad pulses. The two areas are positively correlated, and events with larger areas have a shorter $\Delta t$.}
    \label{fig:muon_pulse_area_time_diff}
\end{figure}

Figure~\ref{fig:persistence} is a persistence plot for double-pulse waveforms collected at millikelvin temperatures.  Figure~\ref{fig:muon_pulse_area_time_diff} shows some analysis of the relationship between the narrow and broad components. Their areas and amplitudes are strongly positively correlated. The delay time between broad and narrow components is also strongly anticorrelated with either component's area and amplitude.  Since both components scale with the amount of detected light, the double pulses are a property of the sensor response to light and not an artifact of the readout electronics. When the light is sufficiently bright, the two components appear to merge. We currently do not have a model to explain how the transition to superconductivity of some material in the SiPMs leads to this phenomenon.

\begin{figure}[h]
    \centering
    \includegraphics[width=1\linewidth]{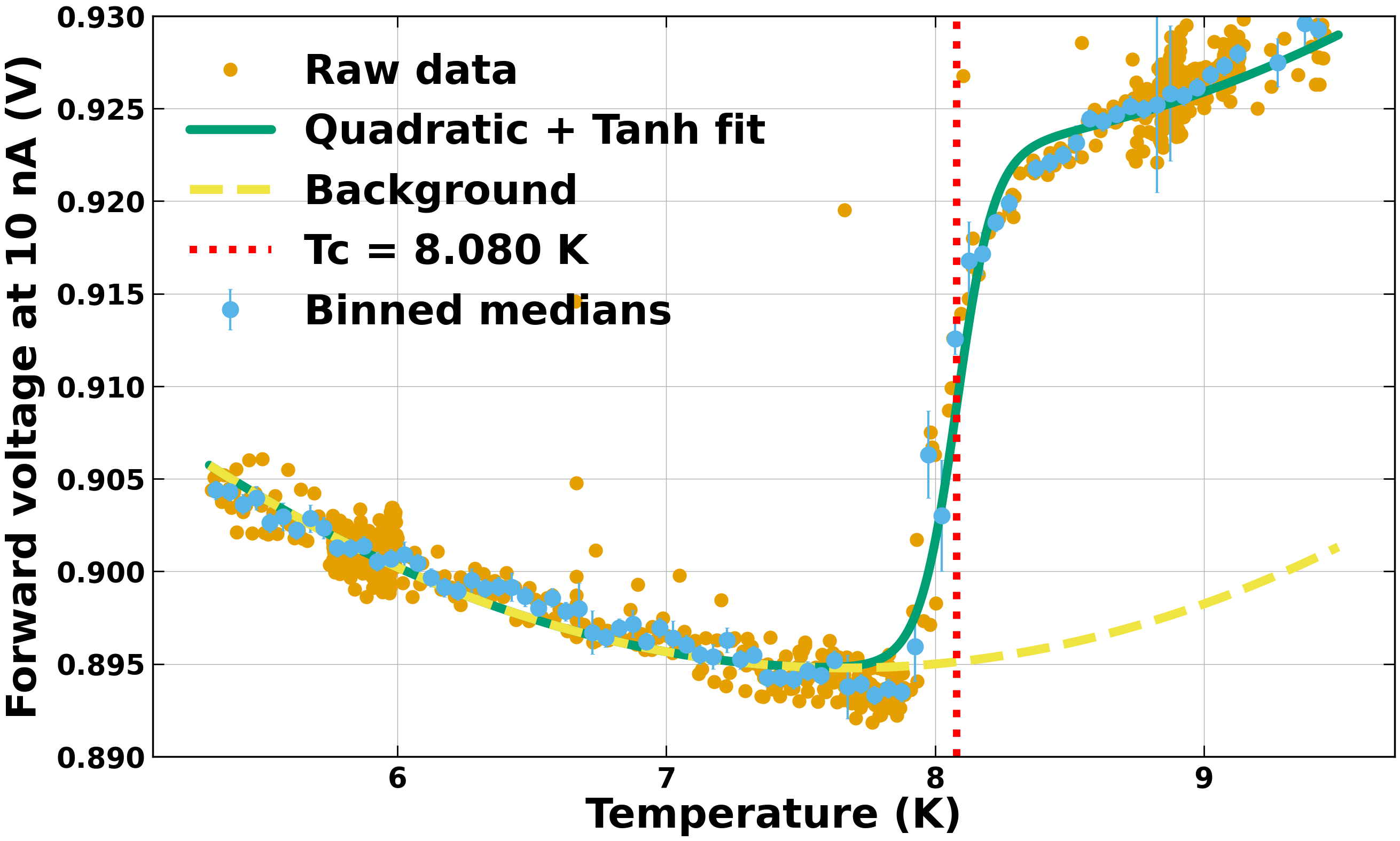}
    \caption{The forward voltage of a SiPM biased at \qty{10}{nA} versus temperature. The plot shows the raw data (orange points), the binned medians (with bin widths of \qty{50}{\milli K}) (blue points), and an empirical transition fit that accounts for the background slope.}
    \label{fig:forwardvoltage}
\end{figure}

\subsection{SiPM voltage drop change with temperature}
To further check for superconductivity, the forward voltage drop of a SiPM biased at a current of \qty{10}{nA} was measured between \qty{5}{K} and \qty{10}{K} in the \emph{Leiden-4K} using a Keithley 2450 SMU under dark conditions. The result is shown in \autoref{fig:forwardvoltage}. The raw data was first binned with a bin width of \qty{50}{\milli K} before being fit to the following empirical transition function which consists of a quadratic part that accounts for the changing baseline of the forward voltage drop and a transition term:
\begin{equation}
    V(T)=a+bT+cT^2+\frac{\Delta V}{2}\left[1+\tanh\left(\frac{T-T_c}{w}\right)\right]
\end{equation}
$T_c$, $w$, and $\Delta V$ are the center, width, and voltage step of the transition respectively. This gives us a transition temperature of \qty{8.0800 \pm 0.0113}{K}, a width of \qty{0.1385 \pm 0.0179}{K}, and a $\Delta V$ of \qty{ 27.655 \pm 0.909}{\milli V}.

\end{document}